\documentclass[preprint2]{aastex}
\shorttitle{High-Mass Cores in the $\eta$ Car GMC}
\shortauthors{Yonekura et al.}

\begin{document}

\title{HIGH-MASS CLOUD CORES IN THE $\eta$ CARINAE GIANT MOLECULAR CLOUD}

\author{\textsc{Yoshinori} \textsc{Yonekura}\altaffilmark{1},
\textsc{Shin'ichiro} \textsc{Asayama}\altaffilmark{1,2},
\textsc{Kimihiro} \textsc{Kimura}\altaffilmark{1},\\
\textsc{Hideo} \textsc{Ogawa}\altaffilmark{1}, 
\textsc{Yoko} \textsc{Kanai}\altaffilmark{3},
\textsc{Nobuyuki} \textsc{Yamaguchi}\altaffilmark{2,3},\\
\textsc{Yasuo} \textsc{Fukui}\altaffilmark{3},
and
\textsc{Peter} J.\ \textsc{Barnes}\altaffilmark{4,5}
}
\altaffiltext{1}{Department of Physical Science, Osaka Prefecture University, 1-1 Gakuen-cho, Sakai, Osaka 599-8531, Japan; yonekura@p.s.osakafu-u.ac.jp}
\altaffiltext{2}{present address: National Astronomical Observatory of Japan, 2-21-1 Osawa, Mitaka, Tokyo 181-8588, Japan}
\altaffiltext{3}{Department of Astrophysics, Nagoya University, Furo-cho, Chikusa-ku, Nagoya 464-8602, Japan}
\altaffiltext{4}{School of Physics A28, University of Sydney, NSW 2006 Australia}
\altaffiltext{5}{School of Physics OMB, University of New South Wales, NSW 2052, Australia}

\begin{abstract}
We carried out an unbiased survey for massive dense cores in the giant molecular cloud associated with $\eta$ Carinae with the NANTEN telescope
in $^{12}$CO, $^{13}$CO, and C$^{18}$O $J$ = 1--0 emission lines.
We identified 15 C$^{18}$O cores, whose typical line width $\Delta V_{\rm comp}$, radius $r$, mass $M$, column density $N$(H$_2$), and average number density $n$(H$_2$) were 3.3 km s$^{-1}$, 2.2 pc, 2.6$\times$10$^3$ $M_\sun$, 1.3$\times$10$^{22}$ cm$^{-2}$, and 1.2$\times$10$^3$ cm$^{-3}$, respectively.
Two of the 15 cores are associated with IRAS point sources whose luminosities are larger than 10$^4$ $L_\sun$, which indicates that massive star formation is occuring within these cores. Five cores including the two with IRAS sources are associated with MSX point sources. We detected H$^{13}$CO$^+$ ($J$ = 1--0) emission toward 4 C$^{18}$O cores, two of which are associated with IRAS and MSX point sources, another one is associated only with an MSX point source, and the other is associated with neither IRAS nor MSX point sources. The core with neither IRAS nor MSX point sources shows the presence of a bipolar molecular outflow in $^{12}$CO ($J$ = 2--1), which indicates that star formation is also occuring in the core, and the other three of the four H$^{13}$CO$^+$ detections show wing-like emission. In total, six C$^{18}$O cores out of 15 (= 40$\%$) are experienced star formation, and at least 2 of 15 (= 13 $\%$) are massive-star forming cores in the $\eta$ Car GMC.
We found that massive star formation occurs preferentially in cores with larger $N$(H$_2$), $M$, $n$(H$_2$), and smaller ratio of $M_{\rm vir}$/$M$.
We also found that the cores in the $\eta$ Car GMC are characterized by large $\Delta V$ and $M_{\rm vir}$/$M$ on average compared to the cores in other GMCs observed with the same telescope.
These properties of the cores may account for the fact that as much as 60--87 \% of the cores do not show any signs of massive star formation.
We investigated the origin of a large amount of turbulence in the $\eta$ Car GMC. We found that turbulence injection from stellar winds, molecular outflows, and supernova remnants which originated from stars formed within the GMC, are not enough to explain the existing turbulence. We propose the possibility that the large turbulence was pre-existing when the GMC was formed, and is now dissipating. Mechanisms such as multiple supernova explosions in the Carina flare supershell may have contributed to form a GMC with a large amount of turbulence.
\end{abstract}

\keywords{ISM: clouds ---ISM: individual(\objectname{$\eta$ Carinae GMC}) ---ISM: molecules ---radio lines: ISM ---stars: formation}

\section{INTRODUCTION}
It is well established that massive stars form in dense cores within giant molecular clouds (GMCs), while low-mass stars form both in dark clouds and GMCs. The formation process of an isolated low-mass star is fairly well understood; the gravitational collapse of a molecular cloud into a protostellar core and the subsequent accretion via a circumstellar disk surrounded by an infalling envelope (e.g., Shu, Adams, \& Lizano 1987 and references therein), followed by the dissipation of momentum into its surroundings through molecular outflows (Lada 1985; Fukui et al.\ 1986, 1993; Fukui 1989; and references therein). The current paradigm is now being tested by observations, and it has become clear that some modifications to the details of protostellar formation are needed (e.g., Andr\'{e} et al.\ 1993; Onishi et al.\ 1996, 1998; Evans 1999 and references therein). 

Concerning massive star formation, an empirical scenario of the later stages of the massive star formation process has been presented (e.g., Kurtz 2000 and references therein); dense ($\sim$ 10$^7$ cm$^{-3}$), compact ($\la$ 0.1 pc), and hot ($\ga$ 100 K) regions called Hot Cores are formed in GMCs, then the UV radiation from newly formed stars within Hot Cores ionizes the surrounding material to produce ultra-compact H~{\scriptsize II} regions (UCHIIs), and finally the central stars become visible, such as the Trapezium cluster in Orion.
On the contrary, evolution before the formation of Hot Cores remains unclear, although some attempts to detect sources at the earliest stages of massive star formation have been made; surveys toward IRAS point sources and maser sources \citep[e.g.,][]{beu02a, mue02, sri02, shi03} have revealed the presence of the dense, compact, and massive cores (density $\ga$ 10$^6$ cm$^{-3}$, temperature $\ga$ 100 K, luminosity $\ga$ 10$^4$ $L_\sun$) sometimes called High-Mass Protostellar Objects \citep[HMPOs;][]{sri02}, some of which are not accompanied by UCHIIs. Although sources without UCHIIs may be younger than the rest, the presence of IRAS point sources and maser sources themselves stongly indicates that stars are already formed within HMPOs.
It is quite natural to suppose that dense, compact, massive, {\it starless} cores (density $\sim$ 10$^6$ cm$^{-3}$, size $\la$ 1.0 pc, mass $\ga$ 10$^3$ $M_\sun$) are the precursors of the HMPOs and Hot Cores \citep[e.g.,][]{eva02}. Systematic studies of these massive {\it starless} cores had not been carried out until recently, due to the difficulty in selecting targets from the existing database. It is to be noted that large-scale surveys for possible candidates for massive {\it starless} cores (density $\sim$ 10$^4$ cm$^{-3}$) have already been made; e.g., \citet{lad91} surveyed the L1630 cloud (Orion B) in CS ($J$ = 2--1).
The importance of the study of {\it starless} cores has been revealed in a series of unbiased surveys for dense cores in low-mass star-forming regions, especially from a viewpoint of setting observational constraints on the evolutional time scales of the theoretical model \citep{oni96, oni98, tac02}. The same is expected for the massive star forming regions and thus unbiased surveys for massive dense cores in massive star-forming regions are also crucial.

One of the most important tasks in the study of massive star formation at the moment is to identify massive dense cores which have not yet experienced  massive-star formation but are capable of future massive-star formation.
High-resolution observations of these cores will give observational constraints on the initial phases of the evolutional model of a massive star; massive stars are formed by direct accretion onto a central protostar through an accretion disk \citep[e.g.,][]{mck03}, which is simple extension of low-mass star formation, or by collisions of intermediate-mass protostars within a young cluster \citep[e.g.,][]{bon98}.
Moreover, such observations are the only method to reveal the initial conditions of massive star forming cores, since the properties of the cores change rapidly soon after the formation of massive stars due to strong UV radiation and stellar winds.
Surveys for Hot Cores and HMPOs toward IRAS point sources and maser sources are, however, biased toward regions which have experienced star formation, and thus it is not possible to obtain information on the initial conditions of the massive star forming cores. Therefore studies of dense cores which have not yet experienced massive-star formation are awaited.
Unbiased surveys for massive dense cores over entire GMCs in molecular lines as well as in the far-infrared and (sub-)millimeter continuum may be the most effective methods to search for massive {\it starless} cores which are capable of future massive-star formation.
Far-IR and submillimeter/millimeter continuum observations are best suited to detect radiation from the dust in {\it starless} cores, since the radiation from cold dust ($T$ $\la$ 20 K) peaks at far-IR ($\gtrsim$ 150 $\mu$m) wavelengths with gradual decrease toward longer wavelengths. Recent development of sensitive instruments, such as bolometer array receivers, makes (sub-)millimeter continuum observations more effective for carrying out a large-scale survey for massive dense cores.
An unbiased survey of entire GMCs with (sub-)millimeter continuum has not yet been made, but several surveys in small fields surrounding IRAS point sources are now in progress for the purpose of detecting massive dense cores located near the IRAS point sources. So far, regions of $\sim$ 10$\arcmin$ $\times$ 10$\arcmin$ around $\sim$ 200 IRAS point sources have been observed at a resolution of 8--24$\arcsec$ \citep[][]{for04, fau04}. Based on these surveys, a few massive dense cold cores without mid-IR (MSX) and far-IR (IRAS) counterparts have been found \citep[][]{for04, gar04}.
Molecular line observations have the invaluable merit of drawing kinematic information within the cores, which is impossible in continuum observations. The empirical relation between the luminosity of the forming star and the line width of the parent dense core has been found by many authors from observations in NH$_3$ \citep[e.g.,][]{sta88, wou88, mye91, har93, lad94, jij99} as well as in C$^{18}$O \citep[e.g.,][]{sai01}, which makes molecular line observations still more important.

We, the NANTEN group, are now conducting unbiased surveys for dense cores in the C$^{18}$O ($J$ = 1--0) line with the NANTEN 4-meter millimeter telescope. \citet{tac02} compiled a sample of 174 C$^{18}$O dense cores in low-mass star-forming regions of Taurus \citep{oni96, oni98}, L1333 \citep{oba98}, Ophiuchus North \citep{noz91}, $\rho$ Oph \citep{tac00}, Chamaeleon \citep{miz99}, Lupus \citep{har99}, Corona Australis \citep{yon99}, the Pipe nebula \citep{oni99}, and the Southern Coalsack \citep{kat99} including intermediate-mass cluster-forming regions such as $\rho$ Oph, Cha I, R CrA, and Lupus 3, and found that cores with active star formation tend to have larger column density, number density, and mass, and tend to be gravitationally more bound. Observations toward massive star forming regions are also in progress; we have detected 112 cores so far in the unbiased surveys for dense cores in GMCs such as Orion A, Orion B, Cepheus OB3, Vela, S35/S37, and Centaurus \citep{yu96, nag97, yam99, sai99, sai01, aoy01}. The beam size of the NANTEN telescope, 2$\farcm$7 ($\sim$0.3 pc at the distance of the nearest GMC, Orion B, $D$ = 400 pc), is not necessarily sufficient to resolve massive {\it starless} cores. However, it is still meaningful to select {\it candidates} for massive {\it starless} cores (density $\sim 10^4$ cm$^{-3}$, size $\sim$ a few pc, mass $\gtrsim$ a few $\times 10^3$ $M_\sun$, which are estimated from the physical parameters of massive {\it starless} cores), which can be traced by C$^{18}$O ($J$ = 1--0) observations with NANTEN, since only a few massive {\it starless} cores have been identified so far. In this paper, we report the results of an unbiased survey for dense cores in the GMC associated with the $\eta$ Carinae nebula (the $\eta$ Car GMC hereafter), with the aim of obtaining samples of massive-{\it starless}-core candidates which are possible sites for future massive star formation.

The $\eta$ Car GMC has been one of the most active star-forming regions in the Galaxy at least until the formation of the youngest stellar cluster associated with the GMC \citep[$\sim$3 Myrs ago, e.g.,][]{fei95}. Eight stellar clusters (Trumpler 14, 15, 16, Collinder 228, Bochum 10, 11, NGC 3293, and NGC 3324; hereafter we abbreviate Trumpler 14 and Trumpler 16 as Tr 14 and Tr 16, respectively) are associated with the GMC, which contains more than 64 O-type stars including one of the most massive stars, $\eta$ Car \citep{fei95}.
When we estimate the star formation activity from the richness of the clusters by counting the number of the most massive stars (e.g., stars with the spectral type of O3) since the total number of the cluster members can not be counted, the $\eta$ Car GMC (with five O3 stars) is estimated to be the second most active star-forming region after NGC 3603 with six O3 stars \citep{mai04}. Its proximity to the Sun ($D$ $\sim$ 2.5 kpc for the $\eta$ Car GMC compared to 6.9 kpc for NGC 3603) makes the $\eta$ Car GMC the best site for the detailed study of massive star formation, because much higher spatial resolution can be achieved.
There is a possibility that active star-forming regions, such as W49, where more than 100 O-type stars have been found in infrared observations \citep{alv03}, are hidden in the galactic plane due to heavy foreground extinction. However, it is natural to think that regions with heavy foreground extinction are located distant from the Sun (11.4 kpc in the case of W49), and thus the $\eta$ Car GMC still remains advantageous. It is to be noted that the well-studied massive star-forming region, the Orion region, is a much smaller system, containing only a single O4-6 star \citep[][]{mai04}.

Several observations have been made so far covering the entire $\eta$ Car GMC with moderate spatial resolution. \citet{gra87, gra88} observed the whole $\eta$ Car GMC (3 $\times$ 2 deg$^2$) in the $^{12}$CO ($J$ = 1--0) line with an 8$\farcm$8 beam, and revealed the existence of a string of CO clouds between $l$ = 284$\fdg$7 and 289$\arcdeg$ with a total mass of 6.7 $\times$ 10$^5$ $M_\sun$. \citet{zha01b} made sub-millimeter observations in $^{12}$CO ($J$ = 4--3) and atomic carbon [CI] $^3P_1$--$^3P_0$ with a 3$\arcmin$ beam, and found that these two emissions and $^{12}$CO ($J$ = 1--0) are approximately coextensive, and that the excitation temperature is higher ($\sim$ 50 K near $\eta$ Car and $\sim$ 20 K in clouds farther from $\eta$ Car) than that in dark clouds.
High-resolution observations at a resolution of $\sim$20--40$\arcsec$ were made toward some selected clouds near $\eta$ Car.
There are two main CO emission regions near $\eta$ Car; northern and southern clouds \citep{deg81, bro98}.
The northern cloud lies surrounding the cluster Tr~14 ($\sim$ 7 pc northwest of $\eta$ Car). \citet{bro03} observed the northern cloud and found that the cloud consists of 4 main CO clouds with masses in the range 40--500 $M_\sun$, two of which are gravitationally bound with density exceeding 10$^5$ cm$^{-3}$. They proposed that these clouds are exposed to a radiation field originating mainly from the most massive stars in Tr 14.
The southern cloud is located adjacent to the cluster Tr 16, which includes $\eta$ Car. \citet{meg96} found a gravitationally-bound cloud with a mass of 70 $M_\sun$, and that a cold IRAS point source with luminosity $\sim$ 10$^4$ $L_\sun$ is associated with the cloud. They concluded that star formation is on-going at the interface between the cloud and the cluster Tr 16, which is triggered by radiation driven shocks.
In between the northern and the southern clouds, the distribution of the molecular gas is patchy due to dissociation by the strong UV radiation and stellar winds from nearby massive stars in Tr 16 \citep{cox95}. They also found that the typical mass of the clouds is $\sim$ 10 $M_\sun$, and that the clouds are not gravitationally bound.
Recently, \citet{rat04} studied the region called the giant pillar \citep{smi00}, $\sim$ 0$\fdg$5 south of $\eta$ Car. Since the distance from the clusters is larger than that of the northern and southern clouds, and the giant pillar is located behind the southren cloud, the influence of the UV radiation and stellar winds from the clusters may be smaller and the photo-dissociation takes longer than the northern and southern clouds.
From CO $J$ = 2--1 observations toward four mid-IR sources identified by \citet{smi00} using mid-infrared data obtained by Midcourse Space Experiment (MSX), they found these sources are also influenced by the radiation from clusters Tr 14 and Tr 16. The mass of the associated clouds ranges from 10--45 $M_\sun$ and they are probably not gravitationally bound. They also found 12 candidates for massive young stellar objects and UCHIIs from the mid-IR data.
These studies revealed that star formation is still on-going in the $\eta$ Car GMC under the strong influence of the most massive stars in clusters, in at least 1.5 $\times$ 2 deg$^2$ regions in the vicinity of $\eta$ Car.

Here we report the results of an unbiased survey for dense cores in the entire $\eta$ Car GMC. The purpose of the present study is to obtain a sample of massive-{\it starless}-core candidates which are possible sites for future massive star formation. With moderate spatial resolution, 2$\farcm$7, our observation covered the entire $\eta$ Car GMC (3 $\times$ 2 deg$^2$) including regions farther from $\eta$ Car, such as the western part of the $\eta$ Car GMC around the H~{\scriptsize II} region Gum 31 and the southern most part of the $\eta$ Car GMC, where no previous observations had been made except coarse large-scale surveys.
Observations were made in C$^{18}$O ($J$ = 1--0) as well as in $^{12}$CO and $^{13}$CO with NANTEN. We also made high-resolution observations in H$^{13}$CO$^+$ ($J$ = 1--0) with the Mopra 22-meter millimeter telescope and searched for molecular outflows in $^{12}$CO ($J$ = 2--1) toward some selected dense cores with the ASTE 10-meter sub-millimeter telescope. Out of the 15 C$^{18}$O cores detected in this survey, two are on-going massive star forming cores associated with luminous infrared sources and another 4 are possibly less-massive star-forming cores, and thus as many as 9--13 cores which have not experienced massive star formation, are detected in this survey. We present the physical properties of the dense cores and compare the star-formation activity with them. We also compare the properties of the C$^{18}$O cores with those in other massive star forming regions, such as Orion, Cepheus, Vela, S35, and Centaurus, in order to study the characteristics of the $\eta$ Car GMC.

%%%%%%%%%%%%%%%%%%%%%%%%%%%%%%%%%
% section 2
%%%%%%%%%%%%%%%%%%%%%%%%%%%%%%%%%
\section{OBSERVATIONS \label{sec_obs}}
\subsection{Large Scale Observations of CO Gas and the Search for High-Mass Cores \label{sec_nanten}}
Observations of the $J$ = 1--0 transitions of $^{12}$CO, $^{13}$CO, and C$^{18}$O were made with the NANTEN 4-meter telescope of Nagoya University at Las Campanas Observatory of the Carnegie Institution of Washington \citep{fuk91, fuk92, fuk98}. The half-power beam width of the telescope was 2$\farcm$7 at 110 GHz and 2$\farcm$6 at 115 GHz, corresponding to $\sim$2 pc for the distance of the $\eta$ Car GMC, 2.5 kpc. The 4 K cooled SIS mixer receiver provided a typical system temperature of $\sim$140 K (SSB) at 110 GHz and $\sim$200 K (SSB) at 115 GHz, including the atmosphere toward the zenith \citep{oga90}. The spectrometer was an acousto-optical spectrometer with a total bandwidth of 40 MHz divided into 2048 channels. The effective spectral resolution was 40 kHz, corresponding to a velocity resolution of 0.11 km s$^{-1}$ at 110 GHz.

The $^{12}$CO data were obtained from 1998 May to July, and in 1999 March. Most of the region was observed by using a position-switching technique with a grid spacing of 2$\arcmin$, and the rest with 4$\arcmin$ grid. The total integration time and the typical rms noise of the data at a velocity resolution of 0.1 km s$^{-1}$ were $\sim$5 sec and $\Delta T_{\rm rms} \sim$ 1.0 K, respectively. In $^{13}$CO, observations were made from 1999 January to March with 4$\arcmin$ grid by using a position-switching technique. The total integration time per point was $\sim$5 sec and the typical rms noise of the data was $\Delta T_{\rm rms} \sim$ 0.4 K at a velocity resolution of 0.1 km s$^{-1}$. C$^{18}$O observations were made in 3 periods from 1998 July to August, in 1999 March, and from 2003 August to October. The data were obtained with a grid spacing of 2$\arcmin$. A frequency-switching technique was used with a switching interval of 13 MHz in 1998 and 1999, whereas a position-switching technique was used in 2003. The total integration time per point was $\sim$90 sec and the typical rms noise of the data was $\Delta T_{\rm rms} \sim$ 0.1 K at a velocity resolution of 0.1 km s$^{-1}$.

For the calibration of the spectral line intensity, a room-temperature chopper wheel was employed. The absolute-intensity calibration was made by observing Ori KL ($\alpha_{1950}$ = 5$^{\rm h}$32$^{\rm m}$47$\fs$0, $\delta_{1950}$=$-$5$\arcdeg$24$\arcmin$21$\arcsec$) and $\rho$ Oph East ($\alpha_{1950}$ = 16$^{\rm h}$29$^{\rm m}$20$\fs$9, $\delta_{1950}$ = $-$24$\arcdeg$22$\arcmin$13$\arcsec$). The peak radiation temperatures, $T_{\rm R}^*$, are assumed to be $T_{\rm R}^*$ ($^{12}$CO, Ori KL) = 65 K \citep{uli76, kut81, lev88}, $T_{\rm R}^*$ ($^{13}$CO, Ori KL) = 10 K \citep{miz95}, and $T_{\rm R}^*$ (C$^{18}$O, $\rho$ Oph East) = 4.4 K. The last value is consistent with the assumption that $T_{\rm R}^*$ (C$^{18}$O, TMC-1 [$\alpha_{1950}$ = 4$^{\rm h}$38$^{\rm m}$42$\fs$0, $\delta_{1950}$ = 25$\arcdeg$35$\arcmin$45$\arcsec$]) = 2.0 K \citep{oni96}, which has been established by the long-term monitoring of standard sources made at the NANTEN telescope. The pointing accuracy was measured to be better than 20$\arcsec$, as checked by optical observations of stars with a CCD camera attached to the telescope as well as by radio observations of Jupiter, Venus, and the edge of the Sun.

%%%%%%%%%%%%%%%%%%%%%%%%%%%%%%%%%
% section 2.2
%%%%%%%%%%%%%%%%%%%%%%%%%%%%%%%%%
\subsection{Detailed Observations for High-Density Regions \label{sec_mopra}}
Observations of the $J$ = 1--0 transition of H$^{13}$CO$^+$ were made with MOPRA, the 22-meter radio telescope of the Australia Telescope National Facility (ATNF), on 2003 July.  The half-power beam width of the telescope was $\sim$37$\arcsec$ at 86 GHz \citep{lad05}. The front end was a dual-channel 4 K cooled SIS mixer receiver, and the typical system temperature was 200 K (SSB) at 86 GHz, including the atmosphere toward the zenith.

Two orthogonal polarizations were simultaneously observed, both tuned to the H$^{13}$CO$^+$ ($J$ = 1--0) frequency. The spectrometer was a digital correlator with a bandwidth of 64 MHz and 2048 channels, which was divided into two 1024 channels in order to measure the two polarizations simultaneously. The velocity coverage was $\sim$ 110 km s$^{-1}$ at 86 GHz, and the effective spectral resolution was $\sim$ 0.13 km s$^{-1}$ at 86 GHz, corresponding to $\sim$ 1.2 times the channel increment.

The data were obtained with a grid spacing of 40$\arcsec$ by the position-switching mode. The total integration time per point was $\sim$ 8 minutes, and the typical rms noise of the data was $\Delta T_{\rm rms} \sim$ 0.1 K at a velocity resolution of 0.1 km s$^{-1}$. The intensity calibration was made by using a room-temperature chopper-wheel. The absolute intensity calibration was made by observing Ori KL, by assuming $T_{\rm R}^*$ = 0.55 K \citep{joh84}.

The pointing accuracy was measured to be within 15$\arcsec$ by observing the SiO maser source RW~Vel at 86 GHz every two hours during the observations.

%%%%%%%%%%%%%%%%%%%%%%%%%%%%%%%%%
% section 2.3
%%%%%%%%%%%%%%%%%%%%%%%%%%%%%%%%%
\subsection{Search for Molecular Outflows \label{sec_aste}}
Observations of the $J$ = 2--1 transition of $^{12}$CO were made with Atacama Submillimeter Telescope Experiment (ASTE), the 10-meter radio telescope of Nobeyama Radio Observatory\footnote{Nobeyama Radio Observatory (NRO) is a branch of the National Astronomical Observatory of Japan (NAOJ), which belongs to the National Institutes of Natural Sciences (NINS)} at Pampa la Bola, Chile, on 2003 November \citep{koh03, eza04}.  The half-power beamwidth of the telescope was 30$\arcsec$ at 230 GHz. The front end was a 4 K cooled SIS mixer receiver \citep{sek01}. The typical system temperature was 300 K (DSB) at 230 GHz, including the atmosphere toward the zenith. We used a digital correlator with a bandwidth of 128 MHz and 1024 channels \citep{sor00}. The effective spectral resolution was 150 kHz, corresponding to a velocity resolution of 0.16 km s$^{-1}$ at 230 GHz.

The data were obtained with a grid spacing of 30$\arcsec$ by using the position-switching mode. The total integration time per point was $\sim$ 1 minute and the typical rms noise of the data was $\Delta T_{\rm rms} \sim$ 0.8 K at a velocity resolution of 0.5 km s$^{-1}$. The intensity calibration was made by using a room-temperature chopper-wheel. The absolute intensity calibration was made by observing Ori KL by assuming $T_{\rm R}^*$ to be 111 K \citep{sut85}. The pointing accuracy was measured to be within 15$\arcsec$ as checked by optical observations of stars with a CCD camera attached to the telescope as well as by radio observations of Jupiter and Mars.

%%%%%%%%%%%%%%%%%%%%%%%%%%%%%%%%%
% section 3
%%%%%%%%%%%%%%%%%%%%%%%%%%%%%%%%%
\section{RESULTS}
\subsection{Distributions of CO Gas \label{sec_cloud}}
The $J$ = 1--0 emission lines of carbon monoxide ($^{12}$CO) and its isotopes ($^{13}$CO and C$^{18}$O) have been used to study molecular clouds, since they are among the most abundant molecules in the dense interstellar medium. Although all these lines have nearly the same critical density for collisional excitation, it is found empirically that the typical densities traced by them are different: $\sim 10^2$ cm$^{-3}$ for $^{12}$CO, $\sim 10^3$ cm$^{-3}$ for $^{13}$CO, and $\sim 10^4$ cm$^{-3}$ for C$^{18}$O (e.g., Mizuno et al.\ 1995, Onishi et al.\ 1996, 1998 for Taurus; Mizuno et al.\ 1998, 1999 for Chamaeleon; Dobashi et al.\ 1994, 1996, Yonekura et al.\ 1997, Kawamura et al.\ 1998, Kim et al.\ 2004 for a series of $^{13}$CO survey in the Milky Way; Tachihara et al.\ 2002 and references therein for C$^{18}$O cores in nearby star forming regions).
This density differentiation can be understood as follows: One of the causes for this density differentiation is due to the optical depth. The $^{12}$CO transition is optically thick in most molecular clouds, making it to probe the lower density enveloping layer of a molecular cloud due to the photon trapping. The other two lines have smaller optical depths, allowing us to look more deeply into clouds. However, because of the selective photo-dissosiation of the $^{13}$CO and C$^{18}$O molecules, abundance ratio of the isotopes vary as a function of the visual extinction $A_{\rm V}$ \citep[e.g., ][]{war96}. The abundance ratio $R$($^{13}$CO/C$^{18}$O) reaches up to $\sim10$ near the cloud surface at $A_{\rm V}$ $\sim$ 1--5. As a result, it is expected that the $^{13}$CO emission mainly comes from the intermediate part of the density ($\sim 10^3$ cm$^{-3}$) and that the C$^{18}$O emission comes from the inner part of the density ($\sim 10^4$ cm$^{-3}$).

\begin{figure}
\epsscale{1}
\plotone{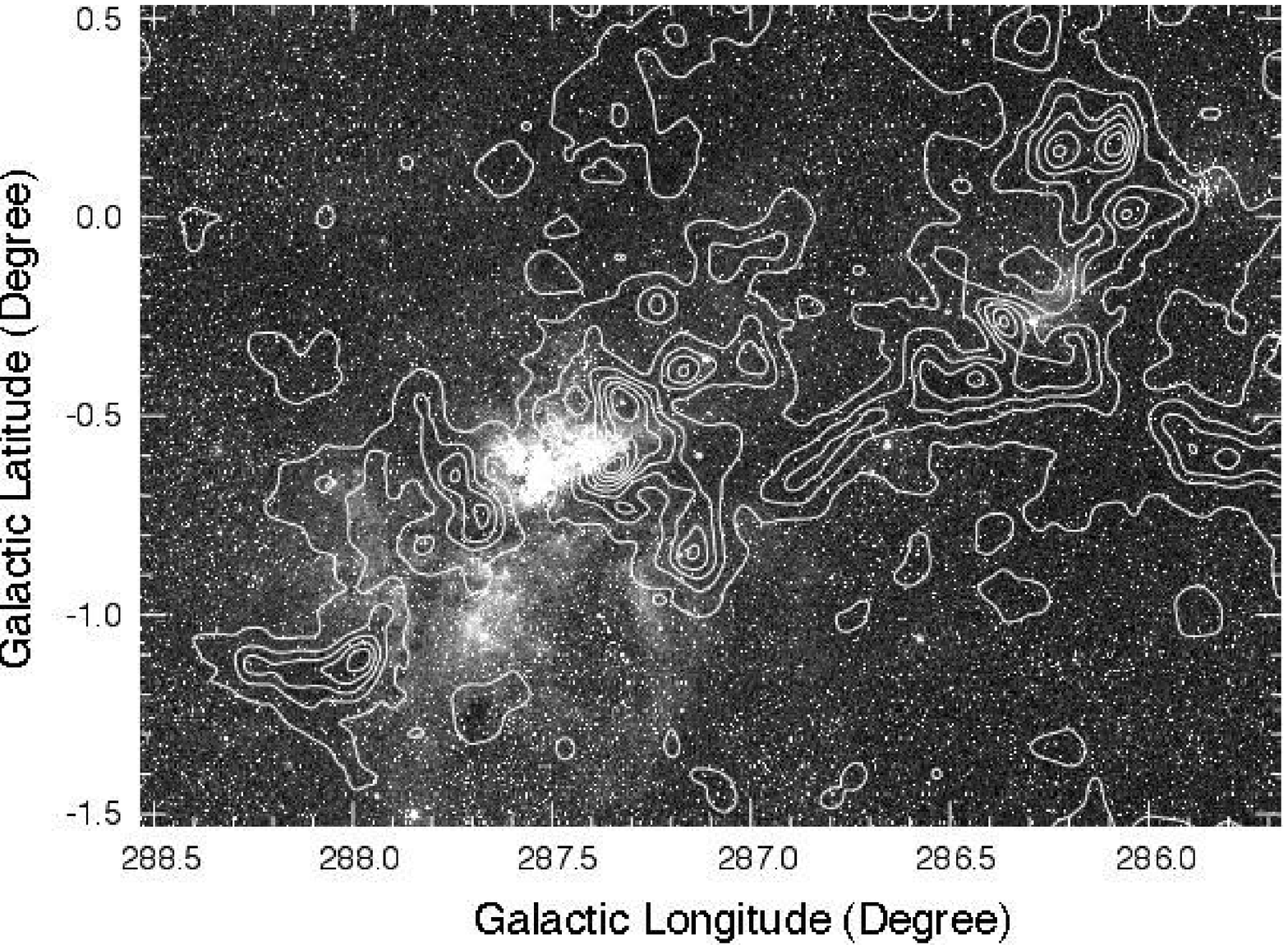}
\caption{Integrated intensity map of the $^{12}$CO ($J$ = 1--0) emission in the velocity range $V_{\rm LSR}$ = $-$30 km s$^{-1}$ to $-$10 km s$^{-1}$, overlaid on the optical image taken from the Digitized Sky Survey. The contour levels are every 20 K km s$^{-1}$, starting from 5 K km s$^{-1}$ (3 sigma).\label{fig01}}
\end{figure}

\begin{figure}
\epsscale{1}
\plotone{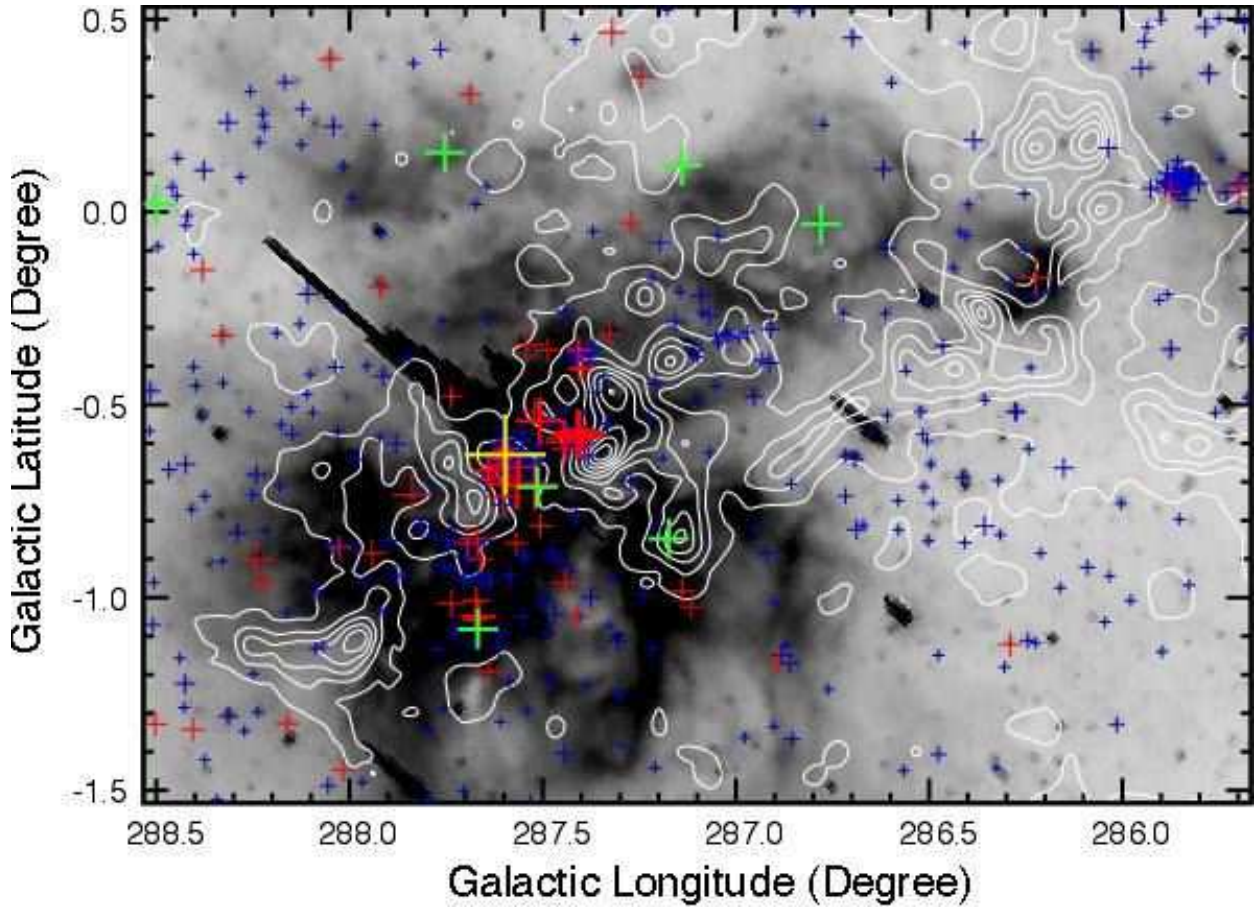}
\caption{Distribution of OB stars is compared with the distribution of $^{12}$CO gas, overlaid on the H$_\alpha$ image reproduced from SHASSA. Yellow, green, red, and blue crosses denote $\eta$ Car, WR stars, O stars, and B stars, respectively. The sizes of the crosses denote the luminosity of the stars. The contour levels are the same as in Figure~\ref{fig01}.\label{fig02}}
\end{figure}

Figure~\ref{fig01} shows the total-intensity distribution of $^{12}$CO ($J$ = 1--0) in the velocity range from $V_{\rm LSR}$ = $-$30 km s$^{-1}$ to $-$10 km s$^{-1}$, overlaid on an optical image taken from the Digitized Sky Survey.
We also show the distribution of OB stars overlaid on a $^{12}$CO map and an H$_\alpha$ image in Figure~\ref{fig02}. The positions of the OB stars are taken from the SIMBAD database. We find $\sim$ 80 O-type stars including five O3 stars, $\sim$400 B stars, and 7 Wolf-Rayet stars in the observed area. The H$_\alpha$ image is reproduced from the Southern H$_\alpha$ Sky Survey Atlas \citep[SHASSA;][]{gau01}.
It is clearly seen that the overall distribution of the $^{12}$CO emission shows an anti-correlation with the optical objects such as stellar clusters and H~{\scriptsize II} regions. Most of the O-type stars are located near the $\eta$ Car nebula at ($l$, $b$) $\sim$ (287.5, $-$0.7), creating the huge H$_\alpha$ emission nebulosity, while small H~{\scriptsize II} regions are made by a few O stars such as NGC 3293 at (285.9, 0.1) and Gum 31 at (286.2, $-$0.2). On the other hand, there is a good correlation of the $^{12}$CO emission with the optical obscuration around the $\eta$ Car nebula; the northern cloud at (287.3, $-$0.7), the southern cloud at (287.7. $-$0.7), and the giant pillar at (288.0, $-$1.1).

In the $^{12}$CO emission, a dozen intense peaks are distributed within more extended diffuse emission. The total molecular mass traced in the $^{12}$CO emission is $\sim$ 3.5 $\times$ 10$^5$ $M_\sun$, which is estimated in the following manner: We assumed an empirical relation that the molecular hydrogen column density, $N$(H$_2$) is proportional to the $^{12}$CO integrated intensity, $I$($^{12}$CO). The conversion factor is called the $X$-factor [$X$ = $N$(H$_2$)/$I$($^{12}$CO)], and the value lies in the range (1--3) $\times$ 10$^{20}$ cm$^{-2}$ (K km s$^{-1}$)$^{-1}$ toward the clouds in the Galactic disk, as estimated by the virial mass, $\gamma$-ray emission, and so on \citep[e.g.,][]{blo86, sol87, str88, ber93}. Here we adopted $X$ = 1.6 $\times$ 10$^{20}$ cm$^{-2}$ (K km s$^{-1}$)$^{-1}$ as derived from observations with the Energetic Gamma-Ray Experiment Telescope \citep[EGRET;][]{hun97}. The total mass of the $^{12}$CO emission is estimated by using
\begin{equation}
M = \mu m_{\rm H} \sum D^2 \Omega N({\rm H}_2 + {\rm He}) = 2.8 m_{\rm H} \sum D^2 \Omega N({\rm H}_2),
\end{equation}
where $D$ is the distance, $\Omega$ is the solid angle subtended by the effective beam size (2$\arcmin$ $\times$ 2$\arcmin$ or 4$\arcmin$ $\times$ 4$\arcmin$), $m_{\rm H}$ is the proton mass, and $\mu$ is the mean molecular weight. The factor of 2.8 is introduced by taking into account the contribution of helium (one He atom for every five H$_2$ molecules), which is the same assumption as $\mu$ = 2.33. The summation is performed over the observed points within the 3-sigma contour level (5 K km s$^{-1}$) of the integrated intensity. 
The distance of the $\eta$ Car GMC is not well determined. \citet{gra88} used $D$ = 2.7 kpc, which is estimated from the distances of associated star clusters. \citet{fei95} summarized the photometric data of the open clusters distributed over the entire $\eta$ Car GMC, and obtained the average value of 2.5 kpc. In the vicinity of $\eta$ Car, values in the range 2.2--2.8 kpc are reported for the distances of Tr 14 and Tr 16 \citep[e.g.,][]{dav97}. In this paper, we adopt the frequently quoted value of 2.5 kpc as the distance of the $\eta$ Car GMC \citep{fei95}. The uncertainty in determining the total mass is estimated to be $\sim 70 \%$ from the equation (1), since the distance $D$ and the conversion factor $X$ include $\sim 10 \%$ and $\sim 50 \%$ uncertainty, respectively.
The total mass obtained in this manner is somewhat smaller than that derived in \citet{gra88}, 6.7 $\times$ 10$^5$ $M_\sun$ (cloud 11 [287.5$-$0.5] in their catalog), who observed the Carina arm with a 1.2-meter telescope at Cerro Tololo. The difference in the total mass basically originates from the different assumptions on the $X$-factor and the distance used in deriving the mass; they used $X$ = 2.8 $\times$ 10$^{20}$ cm$^{-2}$ (K km s$^{-1}$)$^{-1}$ and $D$ = 2.7 kpc.
When we use the same assumption as in \citet{gra88}, 7.2 $\times$ 10$^5$ $M_\sun$ is obtained, which is in good agreement with the value derived by \citet{gra88}.

\begin{figure}
\epsscale{1}
\plotone{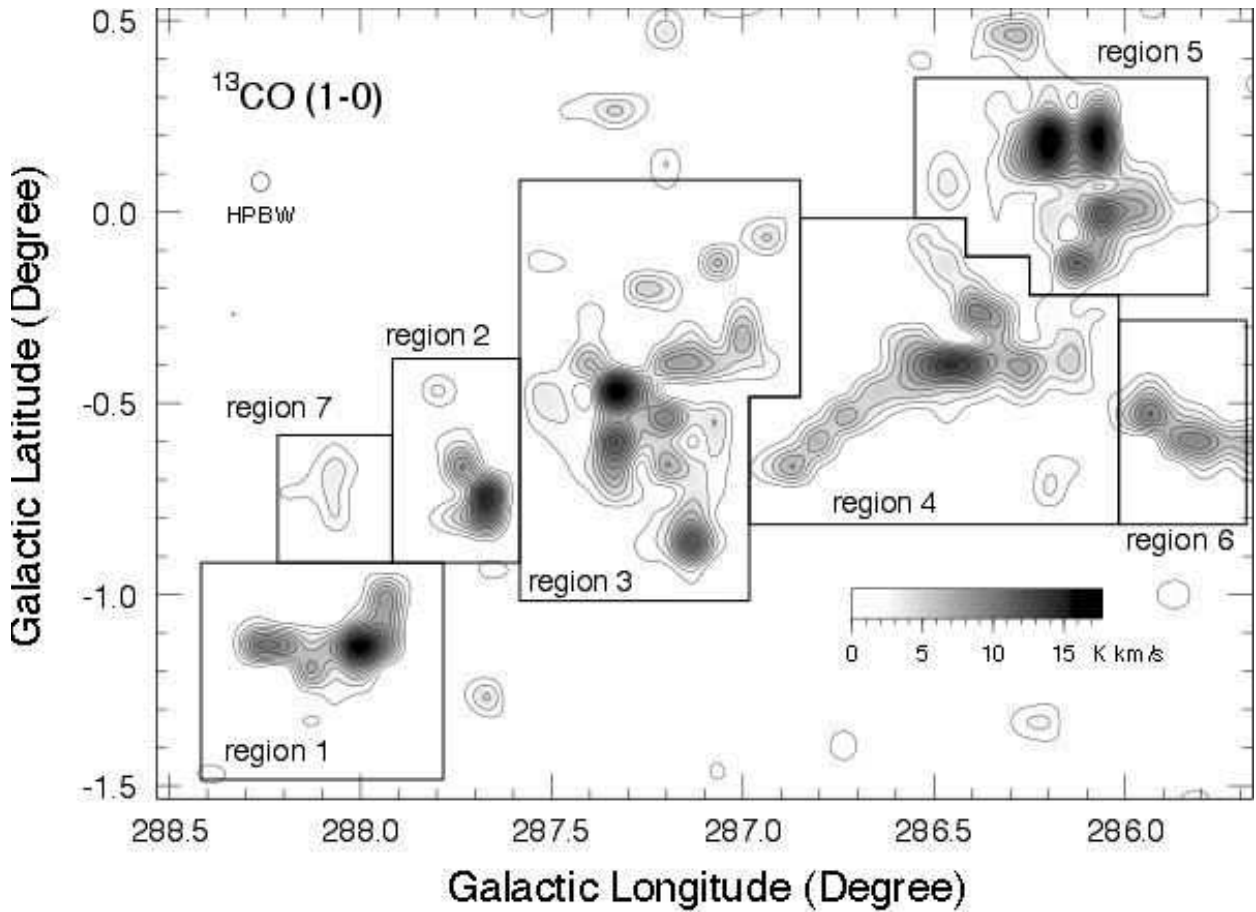}
\caption{Total intensity map of $^{13}$CO ($J$ = 1--0) integrated over the velocity range $V_{\rm LSR}$ = $-$30 km s$^{-1}$ to $-$10 km s$^{-1}$. The contour levels are every 1.35 K km s$^{-1}$ (3 sigma), starting from 1.35 K km s$^{-1}$.\label{fig03}}
\end{figure}

Clumpy structure is more clearly seen in the $^{13}$CO ($J$ = 1--0) emission (Fig.~\ref{fig03}). From the $^{13}$CO distribution, we can divide the $\eta$ Car GMC into 7 regions, which are clearly separated at the 3-sigma contour level (1.35 K km s$^{-1}$). The boundaries of the regions are similar to those in a previous study by \citet{zha01b}, which is based on the $^{12}$CO ($J$ = 1--0) map of 7$\farcm$5 grid spacing with an 8$\farcm$8 beam, but our definition traces more detailed structure. Moreover we identified one more region (region 7). This is mainly because the spatial resolution in our study is $\sim$ 3 times higher than that in \citet{zha01b} and because we used $^{13}$CO emission instead of $^{12}$CO.

The physical properties of the $^{13}$CO gas are estimated in the following manner assuming local thermodynamical equilibrium (LTE): The excitation temperature, $T_{\rm ex}$, was estimated from $T_{\rm R}^*$ ($^{12}$CO) at the peak position of each C$^{18}$O core (see \S~3.2) by using
\begin{equation}
T_{\rm ex} = \frac{5.53}{\ln\{1 + 5.53/[T_{\rm R}^* (^{12}{\rm CO})({\rm K})+0.819] \}} ({\rm K}).
\end{equation}
We used the average value of the $T_{\rm ex}$ of each of the C$^{18}$O cores within the region as the $T_{\rm ex}$ of the region. In order to derive the $^{13}$CO column density, we divide each spectrum into 0.1 km s$^{-1}$ bins, calculate the column density within each bin, and sum them up within an LSR velocity range from $-$30 km s$^{-1}$ to $-$10 km s$^{-1}$. The optical depth in each bin, $\tau_{13}$ ($V$), is calculated by the following equation:
\begin{equation}
\tau_{13} (V) = -\ln\left(1-\frac{T_{13} (V)}{5.29 \{J_{13}[T_{\rm ex} ({\rm K})]-0.164\}}\right),
\end{equation}
where $T_{13}$ ($V$) is the average temperature of the $^{13}$CO spectrum in each bin in the unit of Kelvin, and $J_{13}$[$T$(K)] = 1/{exp[5.29/$T$(K)] $-$ 1}. The $^{13}$CO column density is estimated from
\begin{equation}
N_{13} = 2.42 \times 10^{14} \sum_V { \frac{0.1 ({\rm km \;s}^{-1}) \tau_{13} (V) T_{\rm ex} ({\rm K})}{1-\exp[-5.29/T_{\rm ex} ({\rm K})]} } ({\rm cm}^{-2}).
\end{equation}
The ratio $N$(H$_2$)/$N$($^{13}$CO) is assumed to be 5 $\times$ 10$^5$ \citep{dic78}. The mass is estimated in the same manner as for the $^{12}$CO. The total molecular mass traced in the $^{13}$CO emission is $\sim$ 1.3 $\times$ 10$^5$ $M_\sun$, which is 38 \% of the $^{12}$CO mass.

\begin{figure}
\epsscale{1}
\plotone{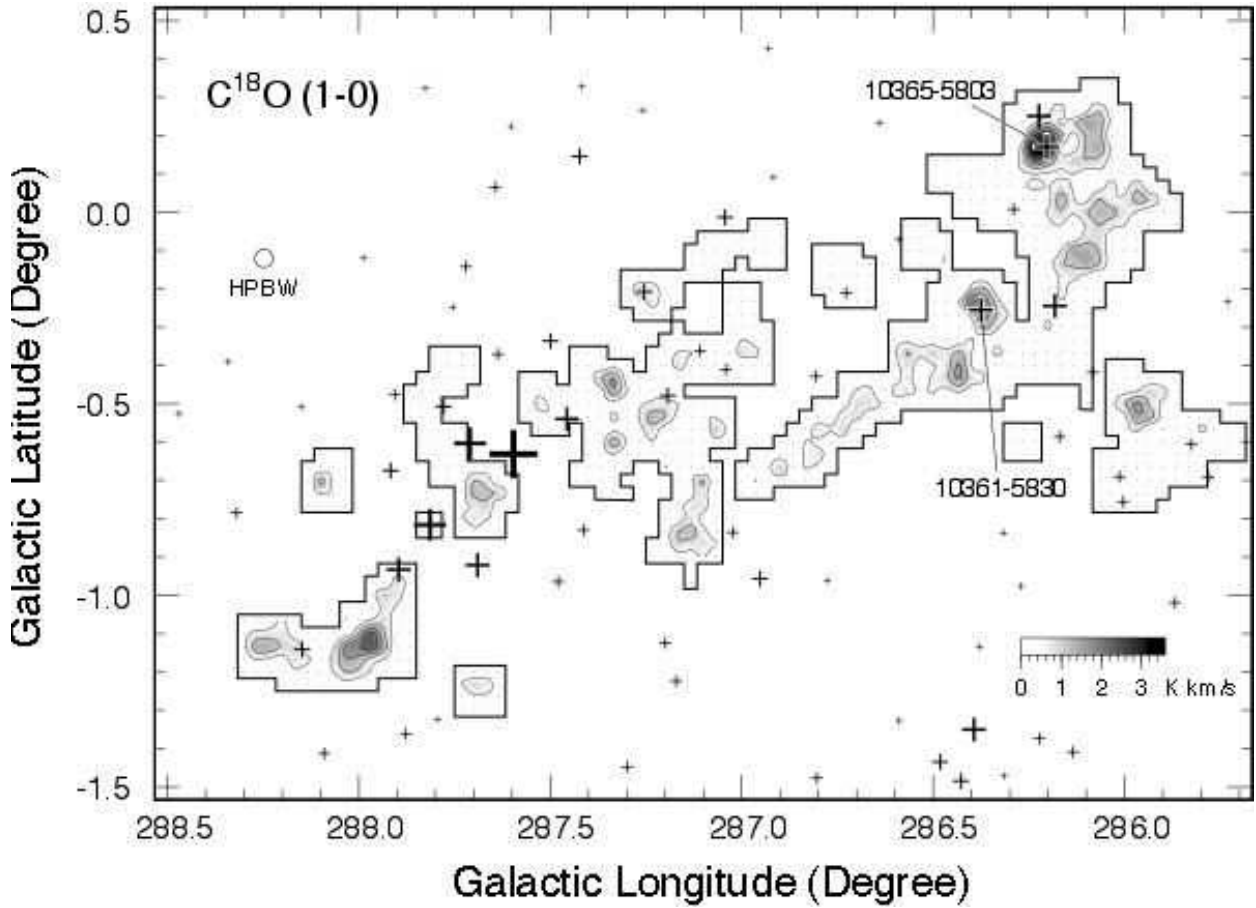}
\caption{Total intensity map of C$^{18}$O ($J$ = 1--0) integrated over the velocity range $V_{\rm LSR}$ = $-$30 km s$^{-1}$ to $-$10 km s$^{-1}$. The contour levels are every 0.5 K km s$^{-1}$ (3 sigma), starting from 0.5 K km s$^{-1}$. IRAS point sources selected as candidates for protostars are also shown as crosses. The sizes of the signs are proportional to the luminosity. Two IRAS point sources which are associated with C$^{18}$O cores are labelled.\label{fig04}}
\end{figure}

The spatial distribution of the C$^{18}$O ($J$ = 1--0) integrated intensity is shown in Figure~\ref{fig04}. The distribution of the C$^{18}$O gas is more compact than those of the $^{12}$CO and $^{13}$CO gas. The total molecular mass traced in the C$^{18}$O emission is $\sim$ 5.8 $\times$ 10$^4$ $M_\sun$ (see \S~3.2), which is 44 \% of the $^{13}$CO mass.

The mass of each region traced in $^{12}$CO, $^{13}$CO, and C$^{18}$O is summarized in Table~\ref{table01}. The fraction of the $^{13}$CO mass to the $^{12}$CO mass ranges from 36--68 \%, with the largest value of 68 \% in region 1. The fraction of the C$^{18}$O mass to the $^{13}$CO mass ranges from 21--72 \%, with the largest value of 72 \% again in regions 1 and 5.

We note that CO emission with $V_{\rm LSR}$ $\ga$ $-$10 km s$^{-1}$ was detected toward some areas around ($l$, $b$) = (286.5, $-$0.2), although these velocity components are not shown in the figures. These velocity components come from distant clouds \citep[clouds on the far side of the Carina arm at $D$ = 7.5 kpc,][]{gra88}, not associated with the $\eta$ Car GMC. We will discuss these clouds in a separate paper (Yonekura et al. 2005 in preparation).

%%%%%%%%%%%%%%%%%%%%%%%%%%%%%%%%%
% section 3.2
%%%%%%%%%%%%%%%%%%%%%%%%%%%%%%%%%
\subsection{Identification of C$^{18}$O Cores and their Physical Properties \label{sec_core}}
In order to investigate the physical properties of the dense regions, we define a C$^{18}$O core in the same manner as that adopted by \citet{oni96}; (1) find a peak-intensity position, (2) draw a contour at the half level of the peak intensity, (3) identify a core unless previously identified cores exist within the half-level contour, (4) find the next intensity peak outside the core, (5) repeat the procedure after (2) until the peak intensity falls down below 6-sigma level (= 1.0 K km s$^{-1}$).

\begin{figure}
\epsscale{1}
\plotone{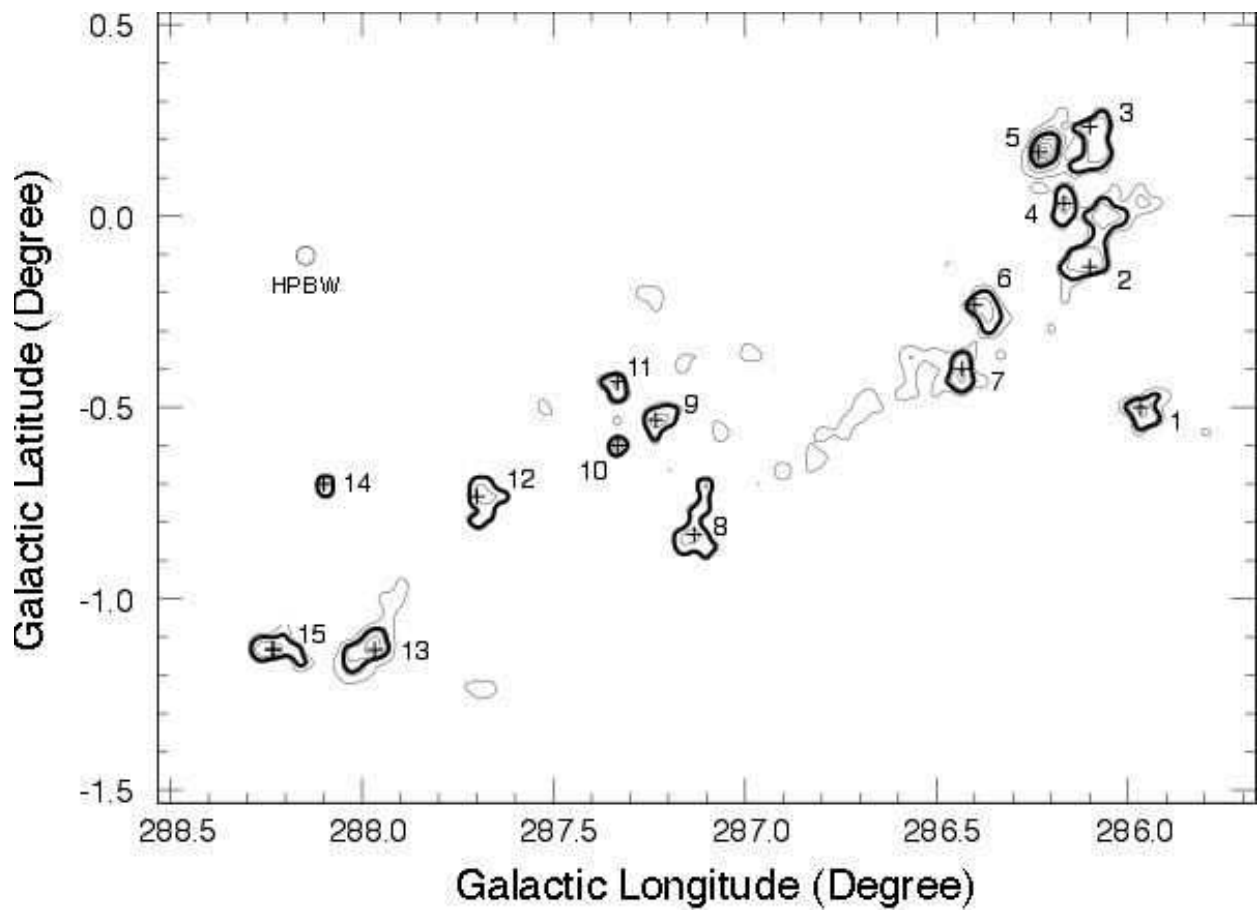}
\caption{Positions of the cores indicated on a C$^{18}$O integrated intensity map. The contour levels are the same as in Figure~\ref{fig04}. Bold contours represent the boundaries of each core.\label{fig05}}
\end{figure}

We finally identified 15 C$^{18}$O cores. In Figure~\ref{fig05} we show the boundaries of each core, and their observed properties are listed in Table~\ref{table02}. The absolute peak temperature, $T_{\rm R}^*$ (C$^{18}$O), the radial velocity of the C$^{18}$O emission, $V_{\rm LSR}$, and the FWHM line width, $\Delta V$, are derived from a single Gaussian fitting for a line profile at the peak position of the core. There is a variation in the ratio of the C$^{18}$O integrated intensity, $I$(C$^{18}$O), to the $^{13}$CO integrated intensity, $I$($^{13}$CO); three prominent C$^{18}$O cores, whose C$^{18}$O intensities at the peak position are larger than $\sim$ 2 K km s$^{-1}$ (nos.\ 5, 6, and 13), possess only moderate $^{13}$CO intensity among the dozens of $^{13}$CO local peaks. This indicates that $^{13}$CO observation is not necessarily suited to trace the high-density regions.

The physical properties of the C$^{18}$O cores were estimated using nearly the same procedure as for $^{13}$CO. Differences between the two procedures are the following: The optical depth of the C$^{18}$O line in each bin, $\tau_{18}(V)$, is calculated by the following equation:
\begin{equation}
\tau_{18} (V) = -\ln\left(1-\frac{T_{18} (V)}{5.27 \{J_{18}[T_{\rm ex} ({\rm K})]-0.166\}}\right),
\end{equation}
where $T_{18}(V)$ is the average temperature of the C$^{18}$O spectrum in each bin in Kelvin, and $J_{18}$[$T$(K)] = 1/{exp[5.27/$T$(K)] $-$ 1}. The C$^{18}$O column density is estimated from
\begin{equation}
N_{18} = 2.42 \times 10^{14} \sum_V { \frac{0.1 ({\rm km \;s}^{-1}) \tau_{18} (V) T_{\rm ex} ({\rm K})}{1-\exp[-5.27/T_{\rm ex} ({\rm K})]} } ({\rm cm}^{-2}).
\end{equation}
The ratio $N$(H$_2$)/$N$(C$^{18}$O) is assumed to be 6 $\times$ 10$^6$ \citep{fre82}.  The mass of a C$^{18}$O core is derived by summing over the observed points within the core. The radius of a core, $r$, was calculated from
\begin{equation}
r = \sqrt{\frac{S}{\pi}},
\end{equation}
where $S$ is the area inside the core. The radius of small cores is overestimated by the beam-size effect. The relation between the observed radius, $r$, and the deconvolved radius, $r_{\rm dec}$, is approximated by the relation
\begin{equation}
r_{\rm dec} = \sqrt{r^2 - \left(\frac{\rm beam\; size}{2}\right)^2}.
\end{equation}
We also calculated the virial mass, $M_{\rm vir}$,
\begin{equation}
\left(\frac{M_{\rm vir}}{M_\sun}\right) = 209 \left(\frac{r}{\rm pc}\right) {\left(\frac{\Delta V_{\rm comp}}{\rm km \; s^{-1}}\right)}^2,
\end{equation}
with $\Delta V_{\rm comp}$ defined as the FWHM line width of the composite profile derived by using a single Gaussian fitting, where the composite profile was obtained by averaging all the spectra within the core. The average H$_2$ number density of the core, $n$(H$_2$), was derived by dividing $M$ by the volume of the core, assuming that the core is spherical with radius $r$. The physical properties of the C$^{18}$O cores thus obtained are listed in Table~\ref{table03} along with the average and the median values.

Here we compare the physical properties of the cores in the $\eta$ Car GMC to those of massive-star-forming cores observed with the NANTEN and the 4-m radio telescopes of Nagoya University: Orion B \citep{aoy01}, Orion A \citep{nag97}, Vela C \citep{yam99}, Cepheus OB3 \citep{yu96}, S35/S37 \citep{sai99}, and Centaurus \citep{sai01}. All these studies assume LTE in deriving physical properties and use the same assumption on abundance. Since the definition of a C$^{18}$O core in Orion A, Vela C, and Cepheus OB3 is different from the one in this work, only values at the peak position (i.e., $\Delta V$ and $N$(H$_2$)) are compared for these regions. In order to be consistent for comparison, we used ($M_{\rm vir}$/$M_\sun$) = 209 ($r$/pc) ($\Delta V$/km s$^{-1}$)$^2$ instead of equation (9), and we excluded cores with serious beam-size effect, i.e., $r_{\rm dec}$/$r$ $<$ 0.8, in calculating the average and median values of $r$, $n$(H$_2$), and $M_{\rm vir}$/$M$ according to \citet{sai01}. Hereafter, we call the cores with $r_{\rm dec}$/$r$ $\ge$ 0.8 as 'resolved' cores.
The average and median values in each region are listed in Table~\ref{table04}. The line width, $\Delta V$, of the cores in the $\eta$ Car GMC ranges from 2.2 to 5.4 km s$^{-1}$ with an average value of 3.2 km s$^{-1}$. This is the largest among the sample. The peak column density ranges from 0.6--3.5 $\times$ 10$^{22}$ cm$^{-2}$ with an average of 1.3 $\times$ 10$^{22}$ cm$^{-2}$, which is a typical value for the GMCs listed here. 
The core mass ranges from 450 to 6100 $M_\sun$ with an average of 2600 $M_\sun$. It is difficult to compare the absolute values of the mass, since there is a dependence on the distance. The radius of the resolved cores ranges from 1.6 to 3.1 pc with an average of 2.2 pc. The radius is also dependent on the distance. The average number density, $n$(H$_2$), ranges from 0.5--3.4 $\times$ 10$^3$ cm$^{-3}$ with an average of 1.2 $\times$ 10$^3$ cm$^{-3}$, which is about one order smaller than the expected value for a C$^{18}$O core, $\sim$ 10$^4$ cm$^{-3}$. This is mainly due to the beam dilution; the resolution of the telescope is $\sim$2 pc at the distance of the $\eta$ Car GMC, which is larger than the typical size of the C$^{18}$O cores in Orion, $\sim$ 0.5 pc. The ratio of the virial mass, $M_{\rm vir}$, to the LTE mass, $M$, ranges from 0.5 to 2.9 with an average of 1.6, which is the second largest value among the whole sample.

To summarize, the cores in the $\eta$ Car GMC are characterized by larger values of $\Delta V$ and $M_{\rm vir}$/$M$ on average than those in other massive-star-forming regions.

%%%%%%%%%%%%%%%%%%%%%%%%%%%%%%%%%
% section 3.3
%%%%%%%%%%%%%%%%%%%%%%%%%%%%%%%%%
\subsection{Star-Formation Activity in Dense Cores: Existence of Large Numbers of Non-Star-Forming Cores \label{sec_iras}}
\begin{figure}
\epsscale{1}
\plotone{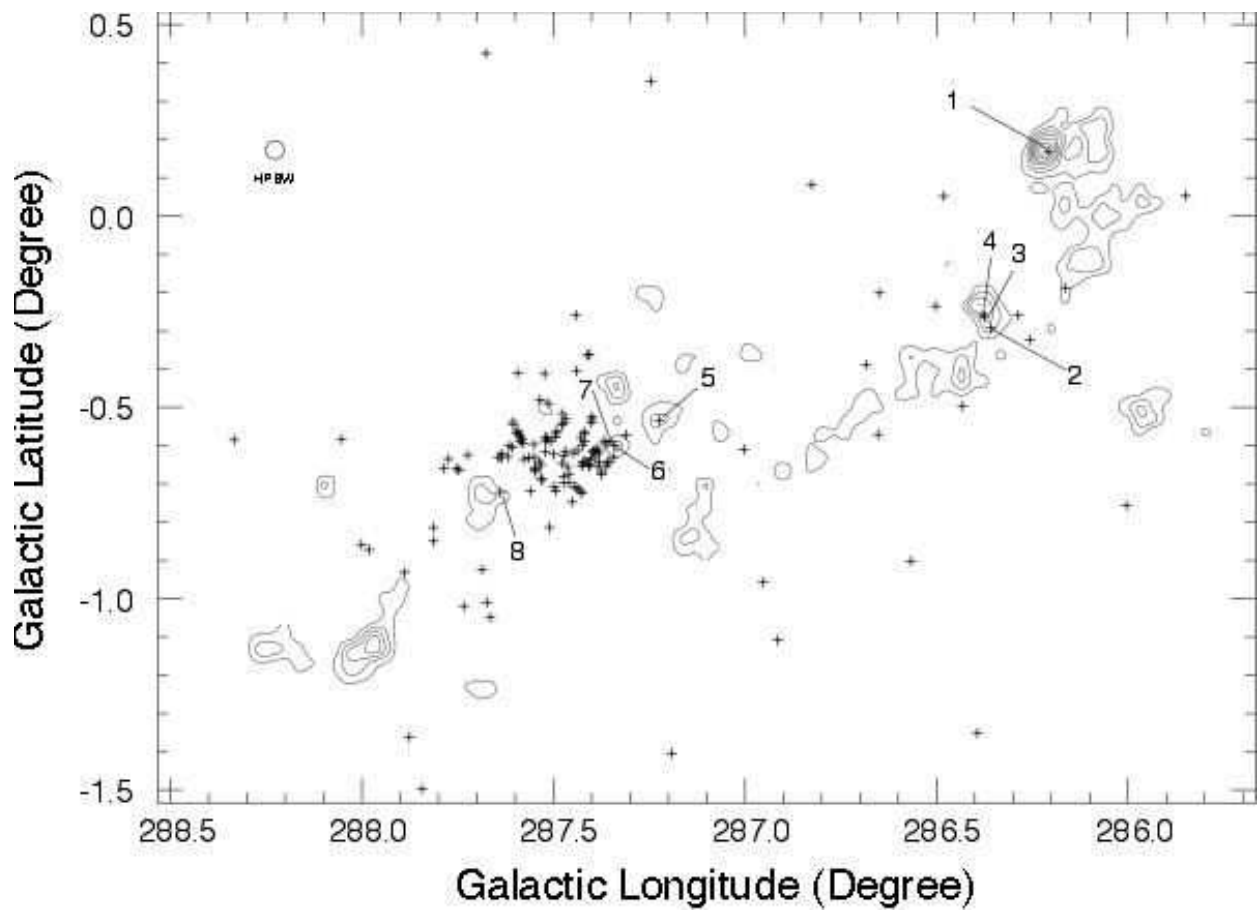}
\caption{MSX point sources selected as candidates for protostars are shown as crosses. Eight MSX point sources which are associated with C$^{18}$O cores are labelled. The contour levels are the same as in Figure~\ref{fig04}.\label{fig06}}
\end{figure}

In this subsection, we investigate the star-formation activity in C$^{18}$O cores in the $\eta$ Car GMC. The IRAS point source catalog is a useful database to search for young stellar objects (YSOs) embedded in molecular clouds. In order to select candidates for recently formed stars among the IRAS point sources, we searched for sources having cold infrared spectra defined at 25 and 60 $\mu$m, i.e., sources with higher flux densities at 60 $\mu$m than at 25 $\mu$m with a data quality better than or equal to 2 in these two bands. We excluded objects from the selection that have already been identified as planetary nebulae or external galaxies in the IRAS point source catalog. Among the 316 IRAS point sources in the observed region, 73 were found to satisfy the criteria, and only two sources are located inside the boundaries of C$^{18}$O cores, i.e., within the half-intensity contours: IRAS 10365$-$5803 and IRAS 10361$-$5830 are associated with C$^{18}$O cores nos.\ 5 and 6, respectively (Figs.~\ref{fig04} and \ref{fig05}). The luminosities of these sources are $\sim$30000 $L_\sun$ (IRAS 10365$-$5803) and $\sim$21000 $L_\sun$ (IRAS 10361$-$5830), indicating that these are candidates for massive stars or star clusters. The properties of these IRAS point sources are summarized in Table~\ref{table05}.
We also searched for YSOs from the MSX point source catalog. We searched for sources with a data quality better than or equal to 3 in Band C ($\sim 12 \mu$m), D ($\sim 14 \mu$m), and E ($\sim 21 \mu$m). Among the 2799 MSX point sources in the observed region, 147 were found to satisfy the criteria, and 8 sources are associated with 5 C$^{18}$O cores (nos.\ 5, 6, 9, 10, and 12; Fig.~\ref{fig06}). The properties of these MSX point sources are summarized in Table~\ref{table06}.

The detection limit of the IRAS survey toward the $\eta$ Car GMC is estimated to be $\sim$ 1000 $L_\sun$, which we derived from the distribution of the luminosity of the above 73 YSO-candidate IRAS point sources on the assumption that these sources are located at the same distance as the $\eta$ Car GMC. Therefore, we can conclude that massive star formation has not occured in the cores without IRAS point sources, although we can not exclude the existence of sources at an evolutionary stage prior to the formation of a massive star, which may not have enough emission at infrared wavelengths \citep[e.g.,][]{sri02}.

Along with the result in the next subsection, 6 C$^{18}$O cores (nos.\ 5, 6, 9, 10, 12, and 15) out of 15 (= 40$\%$) are experienced star formation, and at least 2 of 15 (= 13 $\%$) are massive-star forming cores in the $\eta$ Car GMC, although the proportion of star-forming cores may vary when high-spatial-resolution molecular line observations and more sensitive infrared and (sub)millimeter observations are made.
It should be noted that as many as 13 out of the 15 C$^{18}$O cores, i.e., 87~\%, are not associated with luminous IRAS point sources ($L$ $>$ 10$^3$ $L_\sun$). The proportion, 13~\%, is much lower than in other massive star forming regions at distances within a factor of two of the $\eta$ Car GMC: 28~\% in S35/S37 and 86~\% in Centaurus~III \citep{sai99, sai01}. We will discuss the inactiveness in detail later in \S~4.3

%%%%%%%%%%%%%%%%%%%%%%%%%%%%%%%%%
% section 3.4
%%%%%%%%%%%%%%%%%%%%%%%%%%%%%%%%%
\subsection{Search for High-Density Regions and Molecular Outflows \label{sec_dense}}
In this subsection, we present the results from the search for high-density regions and molecular outflows, in order to reveal the physical properties and the evolutionary stages of dense cores, especially those without IRAS point sources, more precisely.

%%%%%%%%%%%%%%%%%%%%%%%%%%%%%%%%%
% section 3.4.1
%%%%%%%%%%%%%%%%%%%%%%%%%%%%%%%%%
\subsubsection{Identification of H$^{13}$CO$^+$ Cores and their Physical Properties \label{sec_h13cop}}
\begin{figure}
\epsscale{0.58}
\plotone{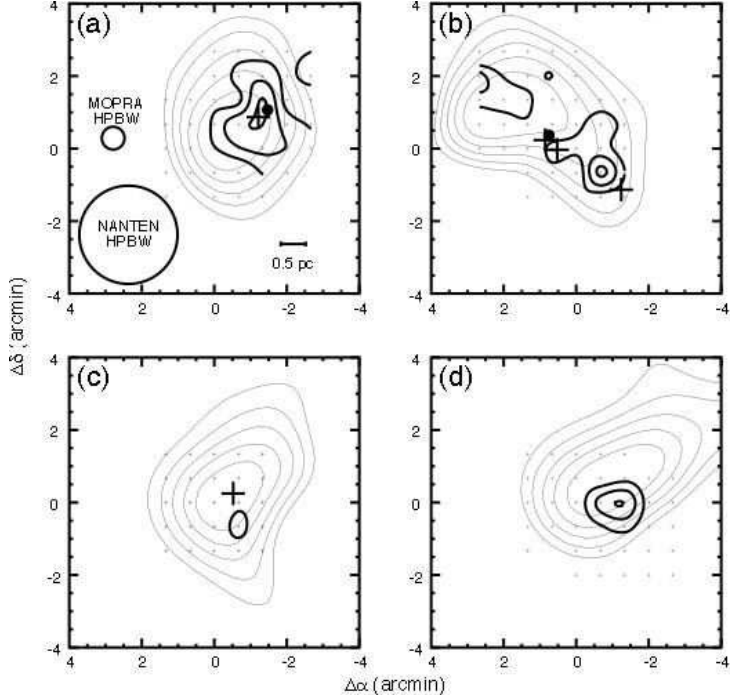}
\caption{(a) Distribution of the H$^{13}$CO$^+$ ($J$ = 1--0) emission toward C$^{18}$O core no.\ 5. The bold contours represent H$^{13}$CO$^+$ intensity integrated over the velocity range $-$23 km s$^{-1}$ $\le$ $V_{\rm LSR}$ $\le$ $-$16 km s$^{-1}$. Contour levels are every 0.4 K km s$^{-1}$ (3 sigma), starting from 0.4 K km s$^{-1}$. The thin contours represent the distribution of the C$^{18}$O core observed by NANTEN, with contour levels of 50 \%, 60 \%, 70 \%, 80 \%, and 90 \% of the peak intensity. The coordinates are shown in offsets from the reference position ($\alpha_{2000}$ = 10$^{\rm h}$ 38$^{\rm m}$ 41$\fs$8, $\delta_{2000}$ = $-$58$\arcdeg$ 20$\arcmin$ 4$\arcsec$). The filled circle and the cross denote IRAS and MSX point sources, respectively.
(b) Same as Fig.\ \ref{fig07}$a$ but for C$^{18}$O core no.\ 6. The velocity range, the 3-sigma level, and the reference position are $-$30 km s$^{-1}$ $\le$ $V_{\rm LSR}$ $\le$ $-$20 km s$^{-1}$, 0.42 K km s$^{-1}$, and (10$^{\rm h}$ 37$^{\rm m}$ 58$\fs$0, $-$58$\arcdeg$ 46$\arcmin$ 40$\arcsec$), respectively.
(c) Same as Fig.\ \ref{fig07}$a$ but for C$^{18}$O core no.\ 9. The velocity range, the 3-sigma level, and the reference position are $-$20 km s$^{-1}$ $\le$ $V_{\rm LSR}$ $\le$ $-$16 km s$^{-1}$, 0.35 K km s$^{-1}$, and (10$^{\rm h}$ 42$^{\rm m}$ 53$\fs$0, $-$59$\arcdeg$ 25$\arcmin$ 44$\arcsec$), respectively.
(d) Same as Fig.\ \ref{fig07}$a$ but for C$^{18}$O core no.\ 15. The velocity range, the 3-sigma level, and the reference position are $-$20 km s$^{-1}$ $\le$ $V_{\rm LSR}$ $\le$ $-$16 km s$^{-1}$, 0.45 K km s$^{-1}$, and (10$^{\rm h}$ 47$^{\rm m}$ 57$\fs$5, $-$60$\arcdeg$ 26$\arcmin$ 26$\arcsec$), respectively.
\label{fig07}}
\end{figure}

In order to detect high-density regions within C$^{18}$O cores, we have observed 14 C$^{18}$O cores (all except for C$^{18}$O core no.\ 4) in H$^{13}$CO$^+$ ($J$ = 1--0) emission, which traces a typical density of $\sim$ 10$^5$ cm$^{-3}$\citep[e.g.,][]{miz94, oni02}. Therefore, the C$^{18}$O cores without IRAS point sources, which are detected in H$^{13}$CO$^+$ emission, can be considered as good candidates for the sites of future massive star formation.
Significant H$^{13}$CO$^+$ emission above the 3-sigma noise level was detected toward 4 C$^{18}$O cores (nos.\ 5, 6, 9, and 15), two of which (nos.\ 5 and 6) have associated IRAS point sources and three of which (nos.\ 5, 6, and 9) with MSX point sources (see $\S$~\ref{sec_iras}). Distributions of H$^{13}$CO$^+$ emission are presented in Figure~\ref{fig07}. In order to study the physical properties, we define the H$^{13}$CO$^+$ cores in the same manner as C$^{18}$O cores. As a result, we identified 3 cores. 

Physical properties of the H$^{13}$CO$^+$ cores were estimated using the same method as for C$^{18}$O: The H$^{13}$CO$^+$ optical depth in each bin, $\tau_V$, is calculated from 
\begin{equation}
\tau ({\rm H}^{13}{\rm CO}^+)_V = -\ln \left(1 - \frac{T({\rm H}^{13}{\rm CO}^+)_V}{4.16 \; [J (T_{\rm ex})-0.272]} \right),
\end{equation}
where $J(T)$ = 1 / [exp(4.16/$T$) $-$ 1]. We tentatively estimate $T_{\rm ex}$ of each H$^{13}$CO$^+$ core from the $^{12}$CO ($J$ = 2--1) observations (see \S~3.4.2), since the beam sizes of both observations are similar. The H$^{13}$CO$^+$ column density, $N$(H$^{13}$CO$^+$), is estimated by using the following equation:
\begin{equation}
N ({\rm H}^{13}{\rm CO}^+) = 1.92 \times 10^{11} \sum_V { \frac{0.1 ({\rm km \;s}^{-1}) \tau({\rm H}^{13}{\rm CO}^+)_V T_{\rm ex} ({\rm K})}{1-\exp[-4.16/T_{\rm ex} ({\rm K})]} } ({\rm cm}^{-2}).
\end{equation}
The H$_2$ column density, $N$(H$_2$), was derived from $N$(H$^{13}$CO$^+$) by making the same assumption as in \citet{aoy01}; [HCO$^+$]/[H$_2$] = 4.0 $\times$ 10$^{-9}$ and $^{12}$C/$^{13}$C = 89.
The assumption of $T_{\rm ex}$ may not be correct, since the density is perhaps smaller than the critical density of H$^{13}$CO$^+$ ($J$ = 1--0), i.e., $J$ = 1--0 line of H$^{13}$CO$^+$ is perhaps sub-thermally excited. However, the uncertainty in $T_{\rm ex}$ would not seriously affect the estimate of the column density $N$: Even if we changed the value of $T_{\rm ex}$ by a factor of 2, the estimated value of $N$ changes within a factor of 1.9, which is smaller than the uncertainty in the HCO$^+$ abundance \citep[e.g., ][]{ber97, jor04}.

The physical properties of the H$^{13}$CO$^+$ cores are listed in Table~\ref{table07}. The total mass of the molecular gas traced by H$^{13}$CO$^+$ in the observed area above the 3-sigma level is estimated to be $\sim$ 4300 $M_\sun$, corresponding to $\sim$ 7 \% of the C$^{18}$O mass.

It is to be noted that a significant difference is found in the behavior of the line width between cores with and without IRAS point sources: The line widths in the H$^{13}$CO$^+$ cores are much smaller than those in the C$^{18}$O cores for the cores without IRAS point sources (nos.\ 9 and 15), whereas the line widths in the H$^{13}$CO$^+$ cores are nearly the same as or somewhat larger than those in C$^{18}$O cores for cores with IRAS point sources (nos.\ 5 and 6).
The present result can be compared to the multitracer, single-cloud line width-size relation, which is classified as $Type~3$ in \citet{goo98}. The relation $\Delta V$ $\propto$ $R^a$ with 0.2 $\la$ $a$ $\la$ 0.7 is reported with the tendency that massive star-forming regions tend to have lower values of $a$ \citep[e.g.,][]{cas95}. The present result is consisitent with the above tendency in a sense that the value $a$ in massive star-forming cores tends to be smaller than in the others, although the trend is more extreme; i.e., $a$ $\la$ 0. The trend might be indicating that line-width enhancement due to the input of turbulence by the activity of star-formation, such as jets and molecular outflows, is effective especially near the center of the massive star-forming cores. However, we have to wait until statistically more accurate studies can be made based on a large sample.

This section can be summarized as follows: 
We detected H$^{13}$CO$^+$ emission from 4 C$^{18}$O cores, which indicates that high-density gas is presented in these cores and that these cores are capable of star formation. Indeed, two (nos.\ 5 and 6) have already experienced massive star formation and another one (no.\ 9) has experienced less-massive star formation. The other C$^{18}$O core (no.\ 15) may have not yet experienced star formation, which makes the core the most probable site for future star formation, though we cannot rule out the possibility that less-massive stars are forming inside (see \S~\ref{sec_outflow}).

%%%%%%%%%%%%%%%%%%%%%%%%%%%%%%%%%
% section 3.4.2
%%%%%%%%%%%%%%%%%%%%%%%%%%%%%%%%%
\subsubsection{Molecular Outflows \label{sec_outflow}}
It is well known that molecular outflows are ubiquitous in low-mass star formation (Lada 1985; Fukui et al.\ 1986; Fukui 1989; Fukui et al.\ 1993 and references therein), and they are believed to be related to the accretion process. It is also suggested by these studies that molecular outflows are also common in massive star forming regions.
Recent systematic surveys for outflows toward regions of massive-star formation showed a detection rate of molecular outflow phenomena as high as $\sim$90 \%, confirming that outflows are also common in massive-star forming regions \citep[e.g.,][]{she96, zha01a, beu02b}.
The formation process of massive stars is, however, still controversial; it is not yet known whether massive stars are formed by direct accretion on a central protostar through an accretion disk \citep[e.g.,][]{mck03}, or by the coalescence of low- to intermediate-mass protostars \citep{bon98}. At any rate, the detection of a molecular outflow is a direct indication of on-going star formation. The age and the luminosity of the driving source can be estimated from the observations, and moreover, the kinematic information, such as the interaction with parent molecular clouds as well as the systemic velocity of the driving source, can be used to solve the distance ambiguity in cases where more than one molecular cloud is located in the same line of sight.

\begin{figure}
\epsscale{1}
\plotone{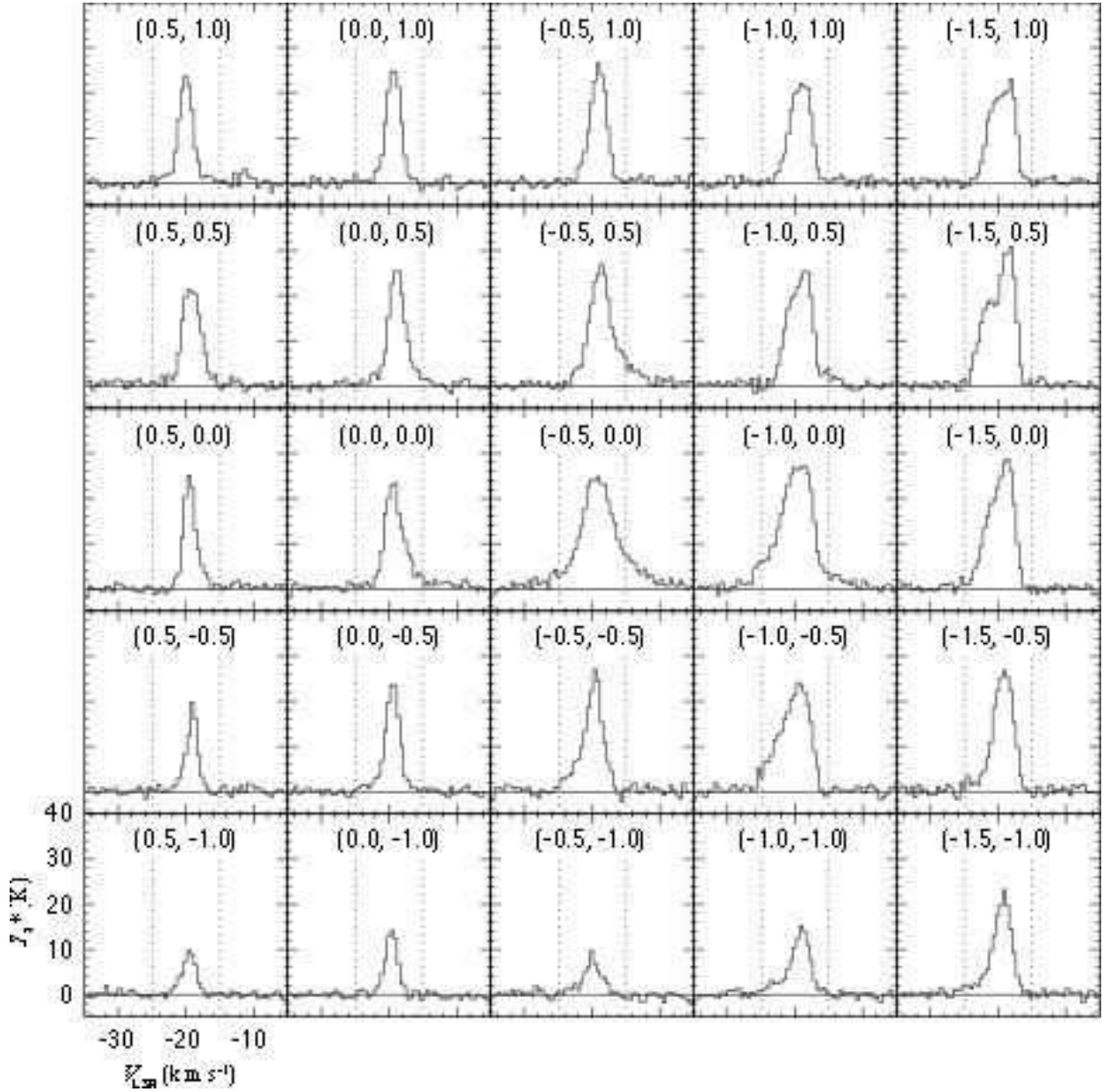}
\caption{$^{12}$CO spectra observed toward C$^{18}$O core no.\ 15. The offsets from the position ($\alpha_{2000}$, $\delta_{2000}$) = (10$^{\rm h}$ 47$^{\rm m}$ 57$\fs$5, $-$60$\arcdeg$ 26$\arcmin$ 26$\arcsec$) are shown in parentheses in the unit of arcminutes. The spectra are smoothed to the velocity resolution of 0.5 km s$^{-1}$.\label{fig08}}
\end{figure}

\begin{figure}
\epsscale{0.9}
\plotone{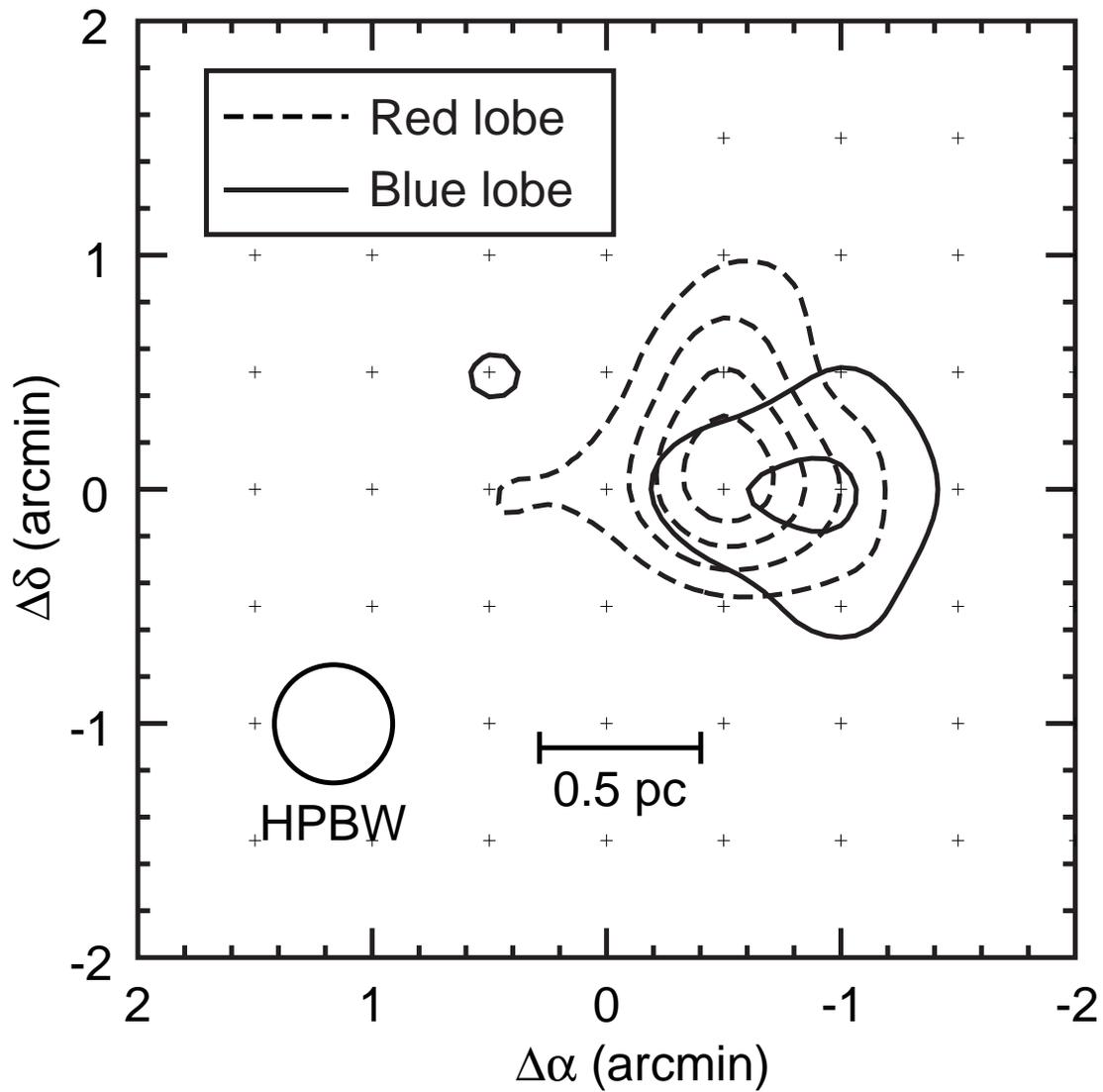}
\caption{Distribution of the molecular outflow within the C$^{18}$O core no.\ 15. The solid contours represent the $^{12}$CO intensity of the blue component integrated between $V_{\rm LSR}$ = $-$35 and $-$25 km s$^{-1}$, and the dashed contours show that of the red component between $V_{\rm LSR}$ = $-$15 and $-$5 km s$^{-1}$. Contours are every 5 K km s$^{-1}$ (3 sigma) from 5 K km s$^{-1}$ as the lowest.\label{fig09}}
\end{figure}

We searched for molecular outflows toward 7 C$^{18}$O cores (nos.\ 5--9, 14, and 15).
We detected one outflow as well as three candidates for outflows. Figure~\ref{fig08} displays the $^{12}$CO spectra around C$^{18}$O core no.~15. In the figure, both the blue-shifted ($-$35 km s$^{-1}$ $\le$ $V_{\rm LSR}$ $\le$ $-$25 km s$^{-1}$) and red-shifted wings ($-$15 km s$^{-1}$ $\le$ $V_{\rm LSR}$ $\le$ $-$5 km s$^{-1}$) are clearly seen in some spectra, where the velocity ranges for wing components are determined at a 1-$\sigma$ noise level in the spectra smoothed to the velocity resolution of 0.5 km s$^{-1}$ ($T_{\rm R}^*$ $\sim$ 0.8 K): The blue-shifted component is prominent at the offset ($\Delta \alpha$, $\Delta \delta$) = ($-$1$\arcmin$, 0$\arcmin$), while the red is most intense at ($-$0$\farcm$5, 0$\arcmin$). The distribution of the $^{12}$CO line intensities integrated between $V_{\rm LSR}$ = $-$35 and $-$25 km s$^{-1}$ (blue component) and between $-$15 and $-$5 (red component) is shown in Figure~\ref{fig09}. 

\begin{figure}
\epsscale{1}
\plotone{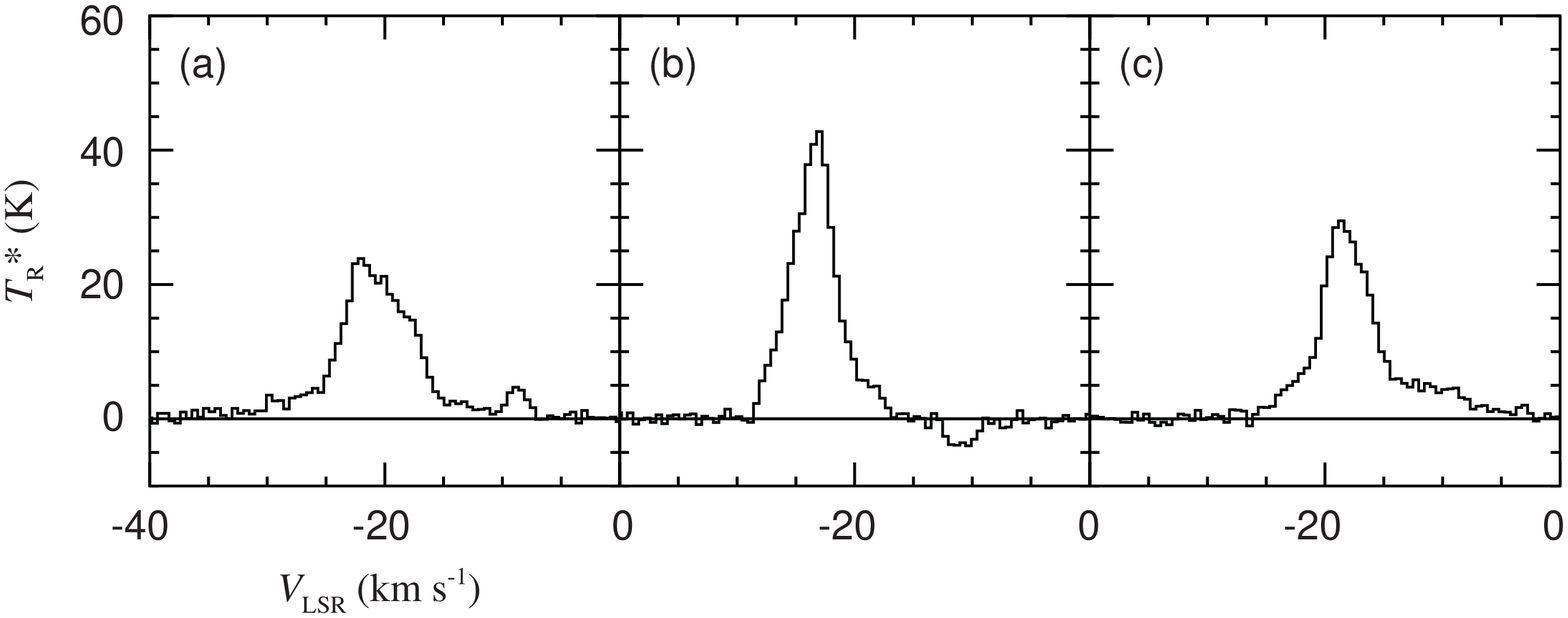}
\caption{$^{12}$CO profiles for the outflow candidates. The positions are (a) $\alpha_{2000}$ = 10$^{\rm h}$ 38$^{\rm m}$ 34$\fs$2, $\delta_{2000}$ = $-$58$\arcdeg$ 19$\arcmin$ 34$\arcsec$ (($\Delta\alpha$, $\Delta\delta$) = ($-$1$\arcmin$, 0$\farcm$5) offset from the reference position in Fig.\ \ref{fig07}$a$), (b) $\alpha_{2000}$ = 10$^{\rm h}$ 37$^{\rm m}$ 54$\fs$1, $\delta_{2000}$ = $-$58$\arcdeg$ 46$\arcmin$ 40$\arcsec$ (($\Delta\alpha$, $\Delta\delta$) = ($-$0$\farcm$5, 0$\arcmin$) offset from the reference position in Fig.\ \ref{fig07}$b$), and (c) $\alpha_{2000}$ = 10$^{\rm h}$ 42$^{\rm m}$ 49$\fs$0, $\delta_{2000}$ = $-$59$\arcdeg$ 25$\arcmin$ 44$\arcsec$ (($\Delta\alpha$, $\Delta\delta$) = ($-$0$\farcm$5, 0$\arcmin$) offset from the reference position in Fig.\ \ref{fig07}$c$).\label{fig10}}
\end{figure}

Figure~\ref{fig10} shows $^{12}$CO profiles of three outflow candidates. Although their spectra show asymmetry, or have broad wing features, we cannot make a contour map of the wing components, due to the existence of multiple velocity components and/or contamination from the emission at OFF position.

Physical parameters of the detected outflow associated with the C$^{18}$O core no.~15 were estimated as follows: The excitation temperature of the high-velocity gas was assumed to be the same as that of the quiescent gas, and estimated to be 30 K from the $^{12}$CO ($J$ = 2--1) observations by using the following equation
\begin{equation}
T_{\rm ex} = \frac{11.1}{\ln\{1 + 11.1/[T_{\rm R}^* (^{12}{\rm CO, \; 2\mbox{\scriptsize --}1})({\rm K})+0.187]\}} ({\rm K}).
\end{equation}
The column densities of  $^{12}$CO ($J$ = 2--1) are calculated under the assumption of LTE using the following equation:
\begin{equation}
N({\rm CO}) = 1.04 \times 10^{13} \; T_{\rm ex} \exp\left(\frac{16.6}{T_{\rm ex}}\right)\frac{1}{\beta}\int T_{\rm R}^* (^{12}{\rm CO})dV,
\end{equation}
where the integration is made over the wing component; $\beta$ is defined from the $^{12}$CO optical depth as $\beta$ = [1 $-$ exp($-\tau$)] / $\tau$. We assumed $\tau$ = 2 for both the red and blue components, which is a typical value for outflows observed in the $^{12}$CO ($J$ = 2--1) line \citep{lev88}.  The column density of H$_2$ molecules was estimated using an [H$_2$]/[CO] ratio of 1 $\times$ 10$^4$ \citep{fre82}. The mass of each component is derived using equation (1) with an effective beam size of 30$\arcsec$ $\times$ 30$\arcsec$. The summation is performed over the observed points within the 3-sigma contour level (5 K km s$^{-1}$) of the intensity integrated over the wing component. Sizes of the lobes are defined at the 3-sigma contour level of the integrated intensity map of the outflow. We tentatively defined the position of the driving source of this outflow as the center of the peak position of each lobe, because no driving sources, such as IRAS and MSX point sources, have been identified around the outflow yet. The radius is defined as the maximum separation between the position of the driving source and the 3-sigma contour of the wing emission.

Using the mass and the radius of the outflow, we estimate the kinetic energy, mechanical luminosity, momentum, and dynamical timescale of the outflow in the same manner as in \citet{yon98}.  We calculated physical properties of the outflow for the velocity ranges from $-$35 to $-$25 km s$^{-1}$ for the blue-shifted wing and from $-$15 to $-$5 km s$^{-1}$ for the red-shifted wing.

These values are summarized in Table~\ref{table08}. The nature of the driving source is not clear because no sources have been identified yet around the outflow. We estimate the upper limit to the luminosity from the IRAS sky survey atlas to be 3900 $L_\sun$. Another estimate for the luminosity can be made by using the empirical relation between the mechanical luminosity of the outflow and the bolometric luminosity of the driving source \citep{cab92}; 15 $L_\sun$ is obtained in this manner.
There are two possibilities for the nature of the driving source; [i] a less-massive star with a luminosity below the IRAS and/or MSX detection limit or [ii] a source at an evolutionary stage prior to the formation of a star without enough emission in the infrared \citep[e.g.,][]{sri02}. We cannot rule out either of the above two possibilities at the moment, but the former might be more probable because the dynamical timescale ($\sim$ 3 $\times$ 10$^4$ yrs), which will probably be overestimated because of the overestimation of size due to low angular resolution, is relatively long. High-resolution CO observations as well as sub-millimeter dust continuum observations will reveal the nature of the driving source.

%%%%%%%%%%%%%%%%%%%%%%%%
% section 4
%%%%%%%%%%%%%%%%%%%%%%%%
\section{DISCUSSION}
\subsection{Physical Properties of C$^{18}$O Cores and their Relation to Star Formation \label{sec_stat}}
We discuss here the relationship between the star-formation activity and the physical properties of C$^{18}$O cores, which has become possible thanks to the detection of a large number of cores without any signs of star formation.

In the study of C$^{18}$O cores in Orion B, \citet{aoy01} found that star formation occurs preferentially in cores with larger $\Delta V$, $N$, and $M$, and that the luminosity of the forming stars increases with $N$. In Centaurus, the luminosity of the forming stars increases with $N$ and $M$ as well as $\Delta V$, and decreases with $M_{\rm vir}$/$M$ \citep{sai01}. \citet{sai01} also found that the luminosity increases with $\Delta V$ by compiling the data in Orion A, B, S35, Vela C, and Centaurus, and proposed a possibility that the mass of forming stars is determined by $\Delta V$, since high mass-accretion rates can be achieved in cores with large $\Delta V$. However, it is controversial whether large $\Delta V$ is a cause or effect of massive star formation.

\begin{figure}
\epsscale{0.65}
\plotone{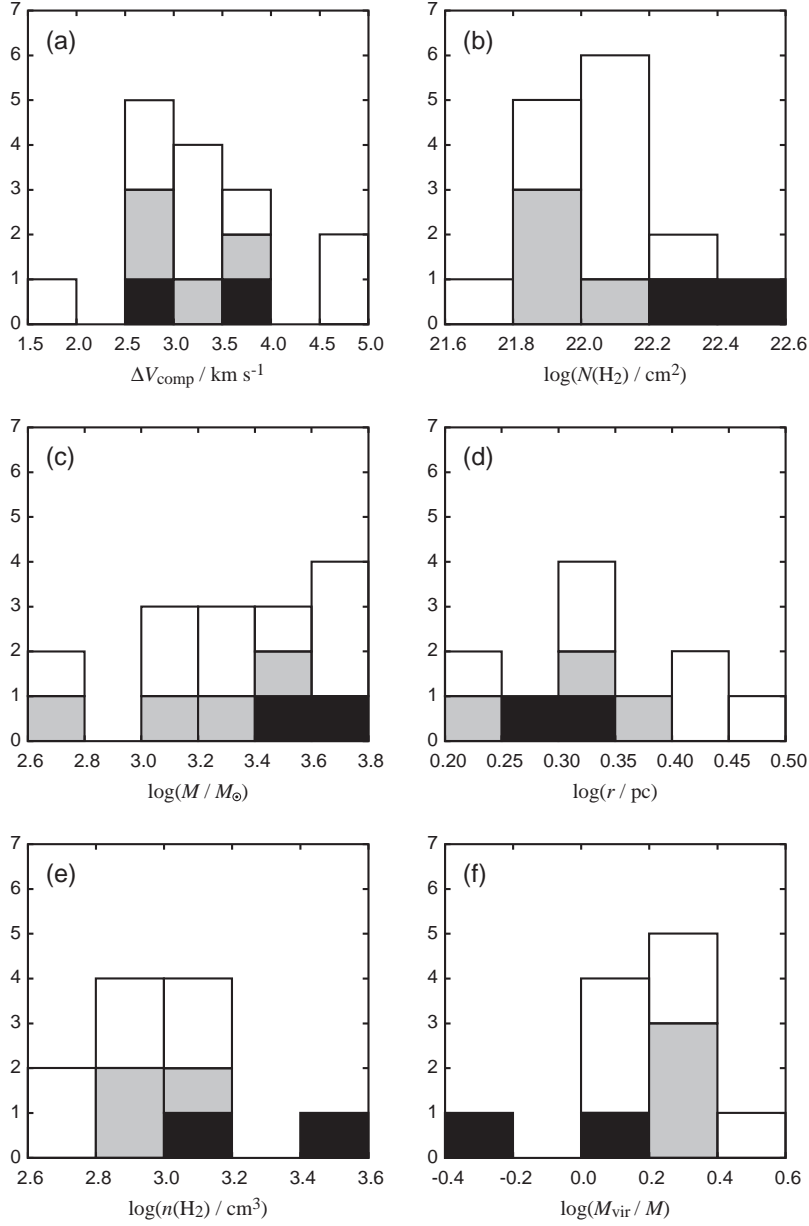}
\caption{Histograms of (a) line width, (b) H$_2$ column density, (c) mass, (d) radius, (e) average H$_2$ number density, and (f) ratio $M_{\rm vir}$/$M$ of cores in the 3 groups. The dark-shaded and light-shaded regions indicate massive-star forming cores (group A) and less-massive-star forming cores (group B), respectively. The open regions indicate less-active star-forming cores (group C). Note that only resolved cores are used to make the histograms of $r$, $n$(H$_2$), and $M_{\rm vir}$/$M$.\label{fig11}}
\end{figure}

We divided the C$^{18}$O cores into 3 groups: (A) cores with IRAS point sources (nos.\ 5 and 6), (B) other star-forming cores (nos.\ 9, 10, 12 and 15), and (C) cores without any signs of star formation. The cores in group A are thought to be sites of massive star formation (see Table~\ref{table05} for the luminosities of the associated IRAS point sources). There are two possibilities for the nature of the cores in group B; these are either [i] less-massive stars with luminosities below the IRAS detection limit or [ii] sources at an evolutionary stage prior to the formation of stars without enough emission in the infrared (see \S~3.4.2). Here we tentatively treat them as less-massive star-forming cores, following the argument in \S~3.4.2.
The cores in group C seem to be sites of less-active star formation than those in groups A and B, which is inferred from the absence of the high-density regions (see \S~3.4.1). Table~\ref{table09} gives the average physical properties of these 3 groups as well as the average of all cores in the $\eta$ Car GMC. We also calculated the average properties of all the star-forming cores, i.e., cores in groups A and B. Figure~\ref{fig11} shows a series of histograms of the core parameters, [a] line width $\Delta V_{\rm comp}$, [b] H$_2$ column density $N$(H$_2$), [c] LTE mass $M$, [d] radius $r$, [e] H$_2$ number density $n$(H$_2$), and [f] ratio $M_{\rm vir}$/$M$. Note that only resolved cores are used to make the histograms of $r$, $n$(H$_2$), and $M_{\rm vir}$/$M$.

It is clearly seen that $N$(H$_2$), $M$, and $n$(H$_2$) are larger and the ratio $M_{\rm vir}$/$M$ is smaller in group A than in groups B and C, while the tendency is not clear in $\Delta V_{\rm comp}$ and $r$ (trend 1).
Between star-forming cores (groups A+B) and non-star-forming cores (group C), $N$(H$_2$) and $n$(H$_2$) are larger in groups A+B than in group C (trend 2).
However, no significant difference between group B and group C is seen (trend 3). 
These trends (trends 1--3) might be indicating the differences in the physical properties among the cores as a function of the masses of the forming stars.

Here we compare the present results to those in Centaurus \citep{sai01}.
In Centaurus, the luminosity of the forming stars increases with $N$ and $M$ as well as $\Delta V$, and decreases with $M_{\rm vir}$/$M$. This trend is same as the trend 1 in $\eta$ Car GMC except $\Delta V$. 
Although we have obtained 127 C$^{18}$O cores in massive star forming regions so far by using the same telescope including 15 in $\eta$ Car GMC, we can not make direct comparison at present, because of the differences in the detection limit and/or the definition of C$^{18}$O cores. The results of detailed comparion will be published in a separate paper.

We can conclude that massive star formation occurs preferentially in cores with larger $N$, $M$, $n$, and smaller $M_{\rm vir}$/$M$ in $\eta$ Car GMC.
High spatial-resolution and high sensitivity observations in infrared (e.g., Spitzer, ASTRO-F) as well as the observations in high density tracer such as CS ($J$ = 7--6), HCO$^+$ ($J$ = 4--3), HCN ($J$ = 4--3) (e.g., ASTE), will help to reveal the properties of C$^{18}$O cores.

%%%%%%%%%%%%%%%%%%%%%%%%%%%%%%%%%
% section 4.2
%%%%%%%%%%%%%%%%%%%%%%%%%%%%%%%%%
\subsection{Comparison with Sub-mm Observations \label{sec_astro}}
It is of interest to compare the distribution of CO and C$^{18}$O ($J$ = 1--0) emissions to that of CO ($J$ = 4--3) emission obtained with the Antarctic Submillimeter Telescope and Remote Observatory \citep[AST/RO;][]{zha01b}, since different excitation conditions are required between CO ($J$ = 4--3) and CO ($J$ = 1--0).
The $J$ = 4 level of the CO molecule is 55 K above the ground level, and the critical density of the $J$ = 4--3 transition is $\sim 8 \times 10^4$ cm$^{-3}$ for the optically thin case, whereas the $J$ = 1 level of the CO molecule is 5.5 K above the ground level, and the critical density of the $J$ = 1--0 transition is $\sim 10^3$ cm$^{-3}$.
At kinetic temperature of 40 K, the $J$ = 4--3 emission of CO is most intense at density of $\sim$ 10$^5$ cm$^{-3}$ while the $J$ = 1--0 intensity is peaked at around 10$^3$ cm$^{-3}$ \citep{gol78}. Observationally, the density effectively emitting these emissions may correspond to somewhat lower than these values ($\sim$ 10$^2$ cm$^{-3}$ for the $J$ = 1--0 emission), which can be understood if we take into account the optical depth effect. Thus, the CO ($J$ = 4--3) line requires higher temperature and density than the CO ($J$ = 1--0) line. However, the overall distribution of the molecular gas traced in the CO ($J$ = 1--0) line and that of the CO ($J$ = 4--3) line is quite similar except the following two regions:

(1) C$^{18}$O core no.~10 (adjacent to Tr 14)\\
The integrated intensities of CO ($J$ = 4--3) and CO ($J$ = 1--0) are the strongest in the observed region, although the integrated intensity of C$^{18}$O ($J$ = 1--0) shows only moderate values. A possible explanation for this is a high excitation temperature due to the star cluster Tr 14: Because of the high excitation temperature, the optical depth of C$^{18}$O ($J$ = 1--0) remains very small in spite of its large column density, resulting in very low intensity in C$^{18}$O ($J$ = 1--0) emission. The NANTEN data indicate the excitation temperature is $\sim$ 25 K, and this value seems to be a lower limit, since higher resolution data obtained with a 24$\arcsec$ beam show the excitation temperature to be as high as $\sim$ 45 K \citep{bro00}.

(2) C$^{18}$O core no.~5\\
The intensity of CO ($J$ = 4--3) is very weak, whereas CO and C$^{18}$O ($J$ = 1--0) emission is quite strong. In general, two explanations are possible; (1) the temperature is not high enough to excite the $J$ = 4--3 emission and/or (2) the density is lower than the typical value traced in CO ($J$ = 4--3). However, the excitation temperature derived from the CO ($J$ = 1--0) observations at NANTEN is 28 K, the highest value among the C$^{18}$O cores in the observed region. The number density, 2800 cm$^{-3}$, is also the highest. Moreover, the detection of H$^{13}$CO$^+$ ($J$ = 1--0) emission toward the core indicates that a high-density region of $>$ 10$^4$ cm$^{-3}$ exists within the core. Thus, the above two hypotheses fail to explain the properties of the core. High-resolution observations of both low-$J$ and mid-$J$ CO lines may unveil the properties of the core.

%%%%%%%%%%%%%%%%%%%%%%%%%%%%%%%%%
% section 4.3
%%%%%%%%%%%%%%%%%%%%%%%%%%%%%%%%%
\subsection{Characteristics of the Star Formation in the $\eta$ Car GMC \label{sec_gmc}}
In this subsection, we will describe the characteristics of the star formation in the $\eta$ Car GMC. The proportion of massive-star forming cores, i.e., cores with IRAS point sources with $L$ $>$ 10$^3$ $L_\sun$, 13 \% (= 2/15), is much lower than in massive star forming regions at distances within a factor of two of the $\eta$ Car GMC; 28 \% in S35/S37 and 86 \% in Centaurus III \citep{sai99, sai01}. This is consistent with the fact that the ratio of $M_{\rm vir}$/$M$ is large (see $\S$~\ref{sec_core}), i.e., most of the cores are not gravitationally bound. The large line widths of the cores ($\sim$ 3 km s$^{-1}$) compared to the other GMCs ($\sim$ 2 km s$^{-1}$, see $\S$~\ref{sec_core}) also support this. A large amount of turbulence in the $\eta$ Car GMC may prevent star formation at the present time.

Let us then consider the origin of the large turbulence in the $\eta$ Car GMC. The kinetic energy due to the turbulence is roughly estimated from $E_{\rm kin}$ = $\frac{1}{2}$ $M$ $V_{\rm turb}^2$. By using the $^{12}$CO mass $M$ = 3.5 $\times$ 10$^5$ $M_\sun$, and the average line width of the C$^{18}$O cores $V_{\rm turb}$ = 3 km s$^{-1}$, $E_{\rm kin}$ $\sim$ 3.4 $\times$ 10$^{49}$ ergs is obtained as the kinematic energy of the $\eta$ Car GMC. The momentum $P$ = $M$ $V_{\rm turb}$ is estimated to be 1.0 $\times$ 10$^6$ $M_\sun$ km s$^{-1}$.
In the absence of any mechanisms to prevent it, the turbulence would dissipate on the crossing time-scale of the cloud, $t_{\rm cr}$ $\sim$ $R_{\rm cloud}$/$\Delta V_{\rm cloud}$ (see V\'{a}zquez-Semadeni et al. 2000 for a review), which corresponds to $\sim$ 15 Myr in the case of the $\eta$ Car GMC. Thus the presence of a large amount of turbulence means either that the turbulence is somewhat constantly injected by some mechanisms, or that a large amount of turbulence is still retained although the turbulence is dissipating because the GMC is young and/or the initial amount of the turbulence is large. In the followings, we examine four possibilities: (1) Turbulence injection by mass outflows from young stellar objects, (2) by supernova remnants, (3) by stellar winds from massive stars, and (4) turbulence is pre-existing.

(1) Turbulence injection by mass outflows from young stellar objects: The kinetic energy and the momentum of such outflows range from $E$ = 10$^{43}$--10$^{47}$ ergs and from $P$ = 0.1--1000 $M_\sun$ km s$^{-1}$, respectively \citep{fuk93}. Thus, even if we assume that all outflows have $E$ = 10$^{47}$ ergs and $P$ = 1000 $M_\sun$ km s$^{-1}$ and that all the energy and momentum of such outflows are converted into turbulence (i.e., conversion coefficient $\varepsilon$ = 1) at least ~1000 YSOs are needed to supply the present turbulence of the $\eta$ Car GMC.

Here we estimate the contribution of outflows from low-mass YSOs, since most of the stars are formed as clusters in massive star forming regions. Observationally, it is known that the kinetic energy and the momentum of outflows are correlated with the bolometric luminosity of the driving source, $L_{\rm bol}$ \citep[][]{wu04}. Least-squares fits to their data yield $P$ = 10$^{-0.48}$ $L_{\rm bol}^{0.59}$ and $E$ = 10$^{-1.87}$ $L_{\rm bol}^{0.68}$. If we assume that the relation between the mass and the luminosity of YSOs as $L$ = $M^a$ with $a$ = 3.45 \citep[][]{all72} and that the initial mass function (IMF) of YSOs as Scalo's IMF \citep[d$n$/d$M$ = $M^{{\rm -}b}$, $b$ = 2.7,][]{sca86} with the cutoff at 1 $M_\sun$, 1.6 $\times$ 10$^5$ YSOs are needed to supply the present turbulence of the $\eta$ Car GMC. The number of YSOs largely depends on the cutoff mass of the IMF and the indices $a$ and $b$, however, 3 $\times$ 10$^3$ are still needed for the IMF with the cutoff at 3 $M_\sun$ with $a$ = 4 and $b$ = 2. Although these estimate are quite rough, the number required to supply the present turbulence of the $\eta$ Car GMC is extremely large compared to the number of YSOs formed in a single GMC in any assumption. Therefore, we can conclude that the turbulence injection by mass outflows from YSOs are not important.

(2) Turbulence injection by supernova remnants: 
The kinetic energy and the initial speed of ejecta of a single supernova remnant are typically $E_{\rm kin}$ = 10$^{51}$ ergs and $V$ = 10$^4$ km s$^{-1}$. If we assume the mass of the ejecta to be $M$ = 3 $M_\sun$, $P$ = 3 $\times$ 10$^4$ $M_\sun$ km s$^{-1}$ can be provided by a single supernova. Thus, even if we assume $\varepsilon$ $\sim$ 1, at least $\sim$30 supernovae are needed to supply the present turbulence of the $\eta$ Car GMC, which is too large when we consider the ages of the associated clusters, 3--7 Myr \citep[e.g.,][]{fei95}. The fact that no supernova remnants are known to exist \citep{whi94} supports this, although possible non-thermal radio emission is reported \citep{jon73, tat91}.
\citet{fuk99} suggested the possible influence of the Carina flare supershell, which may have been formed by more than 20 supernovae located 100 pc away from the $\eta$ Car GMC. The momentum supplied to the $\eta$ Car GMC from the Carina flare supershell through the cross section of the $\eta$ Car GMC is estimated in the following manner: The total momentum of the Carina flare supershell is assumed to be 20 times the typical momentum of a single supernova. We also assume the depth of the GMC to be the same as the projected extent of the GMC, $\sim$ 130 pc. The momentum supplied through the cross section of the $\eta$ Car GMC is estimated to be $P$ = 3 $\times$ 10$^4$ $\times$ 20 $\times$ (130 $\times$ 130)/(4$\pi$ $\times$ 100$^2$) = 8 $\times$ 10$^4$ $M_\sun$ km s$^{-1}$ on the assumption of $\varepsilon$ = 1, and thus it fails to explain the origin of the turbulence.

(3) Turbulence injection by stellar winds from massive stars: The terminal velocity of the stellar wind and the mass-loss rate from $\eta$ Car are $V$ = 700 km s$^{-1}$ and $\dot{M}$ = 5 $\times$ 10$^{-4}$ $M_\sun$ yr$^{-1}$, respectively \citep{boe03, smi03}. During the time scale of 10$^6$ yr, $\eta$ Car will be able to provide $E_{\rm kin}$ = 2 $\times$ 10$^{51}$ ergs and $P$ = 3 $\times$ 10$^5$ $M_\sun$ km s$^{-1}$, which is comparable to the energy and momentum needed if the conversion efficiency is $\sim$ 100 \%. However, the effect of the stellar wind will be confined to a small region around $\eta$ Car; a radius of $\sim$ 25 pc is derived from the models of
\citet{cas75} and \citet{wea77} by taking the typical density of a molecular cloud, 100 cm$^{-3}$, and thus injection from $\eta$ Car alone is not sufficient to explain the observed turbulence.

There are 7 WR stars and $\sim$80 O-type stars around the $\eta$ Car GMC (see Fig.~\ref{fig02}). 
A typical value of the terminal velocity of the stellar wind and the mass-loss rate of a WR star is $V$ $\sim$ 2000 km s$^{-1}$ and $\dot{M}$ = 4 $\times$ 10$^{-5}$ $M_\sun$ yr$^{-1}$ \citep{huc01}. The total energy and momentum from these WR stars in 10$^6$ yr are estimated to be $E_{\rm kin}$ = 1.1 $\times$ 10$^{52}$ ergs and $P$ = 6 $\times$ 10$^5$ $M_\sun$ km s$^{-1}$, respectively.
For an O-type star, typical values are $V$ $\sim$ 1500--2500 km s$^{-1}$ and $\dot{M}$ $\sim$ 10$^{-5}$--10$^{-7}$ $M_\sun$ yr$^{-1}$ \citep[e.g.,][]{mar04}. Over a time scale of 10$^6$ yr, one O-type star will be able to provide $E_{\rm kin}$ $\sim$ 2.3 $\times$ 10$^{48}$--6.3 $\times$ 10$^{50}$ ergs and $P$ $\sim$ 1.5 $\times$ 10$^2$--2.5 $\times$ 10$^4$ $M_\sun$ km s$^{-1}$. Thus about one hundred O stars can provide enough $E_{\rm kin}$ and $P$ on the assumption of $\varepsilon$ = 1. It may therefore be possible that somewhat uniformly distributed WR stars and O-type stars account for the large turbulence, although the estimate of $\varepsilon$ is highly uncertain.

(4) Turbulence is pre-existing:
\citet{ber92} argued that massive cores are magnetically supercritical. This conclusion is extended to low-mass cores by \citet{nak98}. Magnetically supercritical cores are supported by turbulence against the pressure of the surrounding medium $P_{\rm s}$. When $P_{\rm s}$ exceeds a critical value $P_{\rm cr}$, the core cannot be in magnetohydrostatic equilibrium and collapses. \citet{nak98} suggested that dissipation of turbulence is the most important process in reducing $P_{\rm cr}$. The condition $P_{\rm cr}$ $<$ $P_{\rm s}$ can, however, be achieved without dissipating turbulence if $P_{\rm s}$ is large. Thus, star formation in molecular clouds with large turbulence is possible.
Here we investigate the possibility that the $\eta$ Car GMC still retains a large amount of turbulence although that turbulence is dissipating, i.e., the $\eta$ Car GMC is not a relaxed system. There are two possibilities; the $\eta$ Car GMC is young and/or it has an initially large turbulence. The former may be inconsistent with the fact that the ages of the associated clusters are not extremely small, 3--7 Myr \citep[e.g.,][]{fei95}, and thus we concentrate the latter possibility.

Concerning molecular cloud formation, various theories have been proposed so far; cloud formation by spontaneous instabilities, by the compression of diffuse interstellar medium, and by random coalescence of existing clouds (see Elmegreen 1993 for a review). The most realistic scenario may be a combination of the above processes.
The origin of the turbulence in molecular clouds is not yet understood, but possible formation mechanisms of a molecular cloud with large turbulence have been proposed: For example, \citet{koy00} investigated the propagation of a shock wave into the interstellar medium and suggested that the postshock region collapses into a cold layer through thermal instability. Their recent two-dimensional calculation shows that the layer breaks up into small cloudlets, in which a large amount of molecules are formed, in several Myr and that the velocity dispersion of the cloudlets is typically several km s$^{-1}$ \citep{koy02}. The fragments then coalesce and become larger clouds with supersonic velocity dispersions. In the case of the $\eta$ Car GMC, multiple supernova explosions may have contributed to a certain degree; \citet{fuk99} suggested that $\sim$20 supernova explosions may have occurred $\sim$ 100 pc away from the $\eta$ Car GMC over the last 2 $\times$ 10$^7$ yr. These supernova explosions may have triggered the formation of the $\eta$ Car GMC as well as the formation of massive stars within the GMC.

Although the above estimate is quite rough, we suggest the possibility that a large amount of turbulence was supplied when the GMC was formed, and is now dissipating. To identify the formation mechanism for the $\eta$ Car GMC is beyond the scope of the paper, but compression due to supernova explosions in the Carina flare supershell may have played an important role. Turbulent injection from stellar winds, supernova remnants, and outflows originating in the stars formed within the GMC may also contribute, supplying turbulence to some degree.

%%%%%%%%%%%%%%%%%%%%%%%%%%%%
% section 5
%%%%%%%%%%%%%%%%%%%%%%%%%%%%
\section{SUMMARY}
We carried out an unbiased survey for massive dense cores in the giant molecular cloud associated with $\eta$ Carinae with the NANTEN telescope in $^{12}$CO, $^{13}$CO, and C$^{18}$O $J$ = 1--0 emission lines. The purpose of the present study is to obtain a sample of massive-{\it starless}-core candidates which are possible sites for future massive star formation. With moderate spatial resolution, 2$\farcm$7, our observations covered the entire $\eta$ Car GMC (3 $\times$ 2 deg$^2$) including regions farther from $\eta$ Car, where no previous observations had been made except coarse large-scale surveys.
We also made high-resolution observations in H$^{13}$CO$^+$ ($J$ = 1--0) with the MOPRA 22-m millimeter telescope and searched for molecular outflows in $^{12}$CO ($J$ = 2--1) toward selected regions with the ASTE 10-m sub-millimeter telescope.
The main results are summarized as follows:

(1) We identified 15 C$^{18}$O cores, whose typical line width $\Delta V_{\rm comp}$, radius $r$, mass $M$, column density $N$(H$_2$), and average number density $n$(H$_2$) were 3.3 km s$^{-1}$, 2.2 pc, 2.6$\times$10$^3$ $M_\sun$, 1.3$\times$10$^{22}$ cm$^{-2}$, and 1.2$\times$10$^3$ cm$^{-3}$, respectively.

(2) Two of the 15 cores are associated with IRAS point sources whose luminosities are larger than 10$^4$ $L_\sun$, which indicates that massive star formation is occuring within these cores. Five cores including the two with IRAS sources are associated with MSX point sources. We detected H$^{13}$CO$^+$ ($J$ = 1--0) emission toward 4 C$^{18}$O cores, two of which are associated with IRAS and MSX point sources, another one is associated only with an MSX point source, and the other is associated with neither IRAS nor MSX point sources. The core with neither IRAS nor MSX point sources shows the presence of a bipolar molecular outflow in $^{12}$CO ($J$ = 2--1), which indicates that star formation is also occuring in the core, and the other three of the four H$^{13}$CO$^+$ detections show wing-like emission. In total, six C$^{18}$O cores out of 15 (= 40$\%$) are experienced star formation, and at least 2 of 15 (= 13 $\%$) are massive-star forming cores in the $\eta$ Car GMC.

(3) We found that massive star formation occurs preferentially in cores with larger $N$(H$_2$), $M$, $n$(H$_2$), and smaller ratio of $M_{\rm vir}$/$M$.
We also found that the cores in the $\eta$ Car GMC are characterized by large $\Delta V$ and $M_{\rm vir}$/$M$ on average compared to the cores in other GMCs observed with the same telescope.
These properties of the cores may account for the fact that as much as 60--87 \% of the cores do not show any signs of massive star formation.

(4) We investigated the origin of a large amount of turbulence in the $\eta$ Car GMC. We found that turbulence injection from stellar winds, molecular outflows, and supernova remnants which originated from stars formed within the GMC, are not enough to explain the existing turbulence. We propose the possibility that the large turbulence was pre-existing when the GMC was formed, and is now dissipating. Mechanisms such as multiple supernova explosions in the Carina flare supershell may have contributed to form a GMC with a large amount of turbulence.

\acknowledgments
The NANTEN project (southern 4-meter radio telescope) is based on a mutual agreement between Nagoya University and Carnegie Institution of Washington. We greatly appreciate the hospitality of all staff members of the Las Campanas Observatory of the Carnegie Institution of Washington. We also acknowledge that this project could be realized by the contribution from many Japanese public donators and companies.
ASTE (Atacama Submillimeter Telescope Experiment) is a joint project between Japan and Chile. The telescope is operated by the ASTE team including NAOJ, University of Tokyo, Nagoya University, Osaka Prefecture University, and Universidad de Chile. We are grateful to all the members of the ASTE team.
We are grateful to the ATNF for thier hospitality, and to the staff of Mopra Observatory, especially Stuart Robertson, for his help in observations. We are also grateful to Peter Barnes for his help in observations. The Australia Telescope is funded by the Commonwealth of Australia for operation as a National Facility managed by CSIRO.
This work made use of the Southern H-Alpha Sky Survey Atlas (SHASSA), which is supported by the National Science Foundation.
The Digitized Sky Surveys were produced at the Space Telescope Science Institute under U.S.\ Government grant NAGW-2166. The images of these surveys are based on photographic data obtained using the Oschin Schmidt Telescope on Palomar Mountain and the UK Schmidt Telescope. The plates were processed into the present compressed digital form with the permission of these institutions.
We are also grateful to Akiko Kawamura, Yoshiaki Moriguchi, Masanori Nakagawa, and Joanne Dawson for their helpful comments.
This work was financially supported in part by Grants-in-Aid for Scientific Research (KAKENHI) from the Ministry of Education, Culture, Sports, Science, and Technology of Japan (MEXT) and Japan Society for the Promotion of Science (JSPS); Nos. 14102003, 14403001, 15071202, and 15071205.

%%%%%%%%%%%%%%
% References
%%%%%%%%%%%%%%
%\clearpage
%

\clearpage
\onecolumn
%%%%%%%%%%%%%%%%%%%%%%%%%%%%%%%%%%%%%%%%%%%%%%%%%%%%%%%%%%%%%%%%%%
%
% Figure 1
%
%%%%%%%%%%%%%%%%%%%%%%%%%%%%%%%%%%%%%%%%%%%%%%%%%%%%%%%%%%%%%%%%%%
\clearpage

%%%%%%%%%%%%%%%%%%%%%%%%%%%%%%%%%%%%%%%%%%%%%%%%%%%%%%%%%%%%%%%%%%
%
% Figure 2
%
%%%%%%%%%%%%%%%%%%%%%%%%%%%%%%%%%%%%%%%%%%%%%%%%%%%%%%%%%%%%%%%%%%
\clearpage

%%%%%%%%%%%%%%%%%%%%%%%%%%%%%%%%%%%%%%%%%%%%%%%%%%%%%%%%%%%%%%%%%%
%
% Figure 3
%
%%%%%%%%%%%%%%%%%%%%%%%%%%%%%%%%%%%%%%%%%%%%%%%%%%%%%%%%%%%%%%%%%%
\clearpage

%%%%%%%%%%%%%%%%%%%%%%%%%%%%%%%%%%%%%%%%%%%%%%%%%%%%%%%%%%%%%%%%%%
%
% Figure 4
%
%%%%%%%%%%%%%%%%%%%%%%%%%%%%%%%%%%%%%%%%%%%%%%%%%%%%%%%%%%%%%%%%%%
\clearpage

%%%%%%%%%%%%%%%%%%%%%%%%%%%%%%%%%%%%%%%%%%%%%%%%%%%%%%%%%%%%%%%%%%
%
% Figure 5
%
%%%%%%%%%%%%%%%%%%%%%%%%%%%%%%%%%%%%%%%%%%%%%%%%%%%%%%%%%%%%%%%%%%
\clearpage

%%%%%%%%%%%%%%%%%%%%%%%%%%%%%%%%%%%%%%%%%%%%%%%%%%%%%%%%%%%%%%%%%%
%
% Figure 6
%
%%%%%%%%%%%%%%%%%%%%%%%%%%%%%%%%%%%%%%%%%%%%%%%%%%%%%%%%%%%%%%%%%%
\clearpage

%%%%%%%%%%%%%%%%%%%%%%%%%%%%%%%%%%%%%%%%%%%%%%%%%%%%%%%%%%%%%%%%%%
%
% Figure 7
%
%%%%%%%%%%%%%%%%%%%%%%%%%%%%%%%%%%%%%%%%%%%%%%%%%%%%%%%%%%%%%%%%%%
\clearpage

%%%%%%%%%%%%%%%%%%%%%%%%%%%%%%%%%%%%%%%%%%%%%%%%%%%%%%%%%%%%%%%%%%
%
% Figure 8
%
%%%%%%%%%%%%%%%%%%%%%%%%%%%%%%%%%%%%%%%%%%%%%%%%%%%%%%%%%%%%%%%%%%
\clearpage

%%%%%%%%%%%%%%%%%%%%%%%%%%%%%%%%%%%%%%%%%%%%%%%%%%%%%%%%%%%%%%%%%%
%
% Figure 9
%
%%%%%%%%%%%%%%%%%%%%%%%%%%%%%%%%%%%%%%%%%%%%%%%%%%%%%%%%%%%%%%%%%%
\clearpage

%%%%%%%%%%%%%%%%%%%%%%%%%%%%%%%%%%%%%%%%%%%%%%%%%%%%%%%%%%%%%%%%%%
%
% Figure 10
%
%%%%%%%%%%%%%%%%%%%%%%%%%%%%%%%%%%%%%%%%%%%%%%%%%%%%%%%%%%%%%%%%%%
\clearpage

%%%%%%%%%%%%%%%%%%%%%%%%%%%%%%%%%%%%%%%%%%%%%%%%%%%%%%%%%%%%%%%%%%
%
% Figure 11
%
%%%%%%%%%%%%%%%%%%%%%%%%%%%%%%%%%%%%%%%%%%%%%%%%%%%%%%%%%%%%%%%%%%
\clearpage
%
%%%%%%%%%%%%%%%%%%%%%%%%%%%%%%%%%%%%%%%%%%%%%%%%%%%%%%%%%%%%%%%%%%
%
% Table 1
%
%%%%%%%%%%%%%%%%%%%%%%%%%%%%%%%%%%%%%%%%%%%%%%%%%%%%%%%%%%%%%%%%%%
\begin{deluxetable}{rrrrrrrrccrll}
\tabletypesize{\scriptsize}
\rotate
\tablecaption{Comparison between $^{12}$CO, $^{13}$CO, and C$^{18}$O Masses in each Region \label{table01}}
\tablewidth{0pt}
\tablehead{
\colhead{Region} & \colhead{$l_{\rm min}$} & \colhead{$l_{\rm max}$} & \colhead{$b_{\rm min}$} & \colhead{$b_{\rm max}$} & \colhead{$M_{12}$\tablenotemark{a}} & \colhead{$M_{13}$\tablenotemark{a}} & \colhead{$M_{18}$\tablenotemark{a}} & \colhead{$M_{13}$/$M_{12}$} & \colhead{$M_{18}$/$M_{13}$} & \colhead{$T_{\rm ex}$\tablenotemark{b}} & \colhead{C$^{18}$O Cores} & \colhead{Associated Objects}\\
& \colhead{($\arcdeg$)} & \colhead{($\arcdeg$)} & \colhead{($\arcdeg$)} & \colhead{($\arcdeg$)} & & & & & & \colhead{(K)} 
}
\startdata
1 & 287.800 & 288.400 & $-$1.467 & $-$0.933 & 2.2 & 1.5 & 1.07 & 0.68 & 0.72 & 20 & 13, 15 & giant pillar\\
2 & 287.600 & 287.900 & $-$0.900 & $-$0.400 & 1.7 & 0.8 & 0.28 & 0.44 & 0.37 & 20 & 12 & $\eta$ Car, southern cloud\\
3 & 286.867 & 287.567 & $-$1.000 & 0.067 & 9.8 & 3.8 & 0.80 & 0.39 & 0.21 & 23 & 8, 9, 10, 11 & $\eta$ Car, northern cloud\\
4 & 286.033 & 286.967 & $-$0.800 & $-$0.033 & 8.0 & 3.0 & 1.00 & 0.38 & 0.33 & 21 & 6, 7 & Gum 31\\
5 & 285.800 & 286.533 & $-$0.200 & 0.333 & 6.6 & 3.2 & 2.29 & 0.48 & 0.72 & 21 & 2, 3, 4, 5 & Gum 31, NGC 3293\\
6 & 285.700 & 286.000 & $-$0.800 & $-$0.300 & 2.0 & 0.7 & 0.27 & 0.36 & 0.38 & 16 & 1 & \\
7 & 287.933 & 288.200 & $-$0.900 & $-$0.600 & 0.4 & 0.2 & 0.06 & 0.48 & 0.31 & 9 & 14 & \\
total & & & & & 34.6 & 13.2 & 5.78 & 0.38 & 0.44 & & & \\
\enddata
\tablenotetext{a}{$M_{12}$, $M_{13}$, and $M_{18}$ represent $^{12}$CO mass, $^{13}$CO mass, and C$^{18}$O mass, respectively. Units are in 10$^4$ $M_\sun$.}
\tablenotetext{b}{Average value of the $T_{\rm ex}$ of each of the C$^{18}$O cores within the region.}
\end{deluxetable}
\clearpage

%%%%%%%%%%%%%%%%%%%%%%%%%%%%%%%%%%%%%%%%%%%%%%%%%%%%%%%%%%%%%%%%%%
%
% Table 2
%
%%%%%%%%%%%%%%%%%%%%%%%%%%%%%%%%%%%%%%%%%%%%%%%%%%%%%%%%%%%%%%%%%%
\begin{deluxetable}{rrrrrcccrrc}
\tabletypesize{\scriptsize}
\rotate
\tablecaption{Observed Properties of the C$^{18}$O Cores \label{table02}}
\tablewidth{0pt}
\tablehead{
& \multicolumn{4}{c}{Coordinate}\\
\cline{2-5}
\colhead{Name} & \colhead{$l$} & \colhead{$b$} & \colhead{$\alpha$} & \colhead{$\delta$} & \colhead{$T_{\rm R}^*$ (C$^{18}$O)} & \colhead{$V_{\rm LSR}$} & \colhead{$\Delta V$} & \colhead{$I_{18}$\tablenotemark{a}} & \colhead{$I_{13}$\tablenotemark{a}} & \colhead{$I_{18}$/$I_{13}$}\\
& \colhead{($\arcdeg$)} & \colhead{($\arcdeg$)} & \colhead{(J2000)} & \colhead{(J2000)} & \colhead{(K)} & \colhead{(km s$^{-1}$)} & \colhead{(km s$^{-1}$)} & & & 
}
\startdata
1 & 285.967 & $-$0.500 & 10 34 23.6 & $-$58 46 56 & 0.6 & $-$16.6 & 2.4 & 1.6 & 9.5 & 0.17\\
2 & 286.100 & $-$0.133 & 10 36 41.1 & $-$58 31 50 & 0.3 & $-$20.7 & 4.5 & 1.4 & 10.6 & 0.14\\
3 & 286.100 & 0.233 & 10 38 03.5 & $-$58 12 40 & 0.5 & $-$20.6 & 2.6 & 1.4 & 16.5 & 0.09\\
4 & 286.167 & 0.033 & 10 37 45.3 & $-$58 25 06 & 0.4 & $-$21.6 & 2.6 & 1.2 & 4.3 & 0.28\\
5 & 286.233 & 0.167 & 10 38 41.8 & $-$58 20 04 & 1.2 & $-$20.0 & 2.9 & 3.6 & 17.6 & 0.20\\
6 & 286.400 & $-$0.233 & 10 38 19.0 & $-$58 45 54 & 0.5 & $-$24.1 & 3.1 & 1.9 & 9.3 & 0.20\\
7 & 286.433 & $-$0.400 & 10 37 54.6 & $-$58 55 36 & 0.6 & $-$13.0 & 2.5 & 1.5 & 13.4 & 0.11\\
8 & 287.133 & $-$0.833 & 10 41 03.7 & $-$59 38 42 & 0.3 & $-$19.9 & 3.8 & 1.2 & 12.2 & 0.10\\
9 & 287.233 & $-$0.533 & 10 42 53.0 & $-$59 25 44 & 0.5 & $-$17.6 & 2.6 & 1.2 & 9.4 & 0.13\\
10 & 287.333 & $-$0.600 & 10 43 19.7 & $-$59 32 05 & 0.3 & $-$18.7 & 4.0 & 1.2 & 12.7 & 0.10\\
11 & 287.333 & $-$0.433 & 10 43 56.8 & $-$59 23 16 & 0.3 & $-$19.3 & 5.4 & 1.5 & 17.1 & 0.09\\
12 & 287.700 & $-$0.733 & 10 45 24.0 & $-$59 49 28 & 0.3 & $-$25.8 & 3.9 & 1.1 & 14.1 & 0.08\\
13 & 287.967 & $-$1.133 & 10 45 48.2 & $-$60 18 09 & 0.7 & $-$21.3 & 2.8 & 2.3 & 17.2 & 0.13\\
14 & 288.100 & $-$0.700 & 10 48 21.3 & $-$59 58 44 & 0.5 & $-$12.1 & 2.2 & 1.1 & 3.4 & 0.32\\
15 & 288.233 & $-$1.133 & 10 47 43.0 & $-$60 25 31 & 0.4 & $-$17.8 & 3.5 & 1.3 & 9.7 & 0.14\\
\enddata
\tablecomments{Units of right ascension are hours, minutes, and seconds, and units of declination are degrees, arcminutes, and arcseconds.}
\tablenotetext{a}{$I_{18}$ and $I_{13}$ represent the integrated intensity of the C$^{18}$O and $^{13}$CO emission, respectively. Units are in K km s$^{-1}$.}
\end{deluxetable}
\clearpage

%%%%%%%%%%%%%%%%%%%%%%%%%%%%%%%%%%%%%%%%%%%%%%%%%%%%%%%%%%%%%%%%%%
%
% Table 3
%
%%%%%%%%%%%%%%%%%%%%%%%%%%%%%%%%%%%%%%%%%%%%%%%%%%%%%%%%%%%%%%%%%%
\begin{deluxetable}{rrccrrrrcrcccc}
\tabletypesize{\scriptsize}
\rotate
\tablecaption{Physical Properties of the C$^{18}$O Cores \label{table03}}
\tablewidth{0pt}
\tablehead{
\colhead{Name} & \colhead{$T_{\rm ex}$} & \colhead{$N$(H$_2$)\tablenotemark{a}} & \colhead{$\langle N(H_2)\rangle$} & \colhead{$r$} & \colhead{$r_{\rm dec}$\tablenotemark{b}} & \colhead{$M$} & \colhead{$n$(H$_2$)} & \colhead{$\Delta V_{\rm comp}$} & \colhead{$M_{\rm vir}$} & \colhead{$M_{\rm vir}$/$M$} & \colhead{IRAS\tablenotemark{c}} & \colhead{MSX\tablenotemark{d}} & \colhead{H$^{13}$CO$^+$\tablenotemark{e}}\\
 & \colhead{(K)} & \colhead{(10$^{21}$ cm$^{-2}$)} & \colhead{(10$^{21}$ cm$^{-2}$)} & \colhead{(pc)} & \colhead{(pc)} & \colhead{($M_\sun$)} & \colhead{(cm$^{-3}$)} & \colhead{(km s$^{-1}$)} & \colhead{($M_\sun$)} & 
}
\startdata
1 & 16 & 10.6 & \phn7.4 & 2.0 & 1.8 & 2100 & 900 & 2.7 & 3000 & 1.4 & \nodata & \nodata & \nodata\\
2 & 15 & \phn9.2 & \phn6.9 & 3.1 & 2.9 & 4600 & 540 & 3.2 & 6500 & 1.4 & \nodata & \nodata & \nodata \\
3 & 21 & 11.1 & \phn8.4 & 2.7 & 2.5 & 4400 & 750 & 3.3 & 6100 & 1.4 & \nodata & \nodata & \nodata \\
4 & 20 & \phn9.1 & \phn7.3 & 1.4 & 1.0 & 1000 & 1200 & 2.6 & 1900 & 1.9 & \nodata & \nodata & \nodata \\
5 & 28 & 34.8 & 25.7 & 1.8 & 1.6 & 6100 & 3400 & 2.8 & 2900 & 0.5 & 1 & 1 & 1 \\
6 & 26 & 17.3 & 12.8 & 2.0 & 1.8 & 3700 & 1600 & 3.6 & 5300 & 1.5 & 1 & 3 & 2 \\
7 & 17 & 10.7 & \phn8.6 & 1.6 & 1.3 & 1600 & 1300 & 3.1 & 3200 & 2.0 & \nodata & \nodata & \nodata \\
8 & 24 & 10.1 & \phn7.0 & 2.7 & 2.5 & 3700 & 630 & 3.6 & 7500 & 2.0 & \nodata & \nodata & \nodata \\
9 & 21 & \phn9.7 & \phn7.3 & 1.6 & 1.3 & 1400 & 1100 & 2.5 & 2200 & 1.6 & \nodata & 1 & 3 \\
10 & 25 & 11.0 & 11.0 & 0.8 & \nodata & 520 & 3300 & 4.0 & 2700 & 5.2 & \nodata & 2 & \nodata \\
11 & 22 & 11.8 & 10.1 & 1.4 & 1.0 & 1400 & 1700 & 4.5 & 6100 & 4.3 & \nodata & \nodata & \nodata \\
12 & 20 & \phn8.8 & \phn6.5 & 2.3 & 2.1 & 2600 & 710 & 3.5 & 5900 & 2.3 & \nodata & 1 & \nodata \\
13 & 20 & 17.4 & 12.8 & 2.2 & 1.9 & 4200 & 1400 & 4.9 & 11000 & 2.6 & \nodata & \nodata & \nodata \\
14 & 9 & \phn5.9 & \phn4.8 & 1.2 & 0.6 & 450 & 1000 & 1.9 & 900 & 2.0 & \nodata & \nodata & \nodata \\
15 & 19 & \phn9.9 & \phn7.0 & 2.0 & 1.8 & 2000 & 850 & 2.9 & 3500 & 1.8 & \nodata & \nodata & 4 \\
average & 20 & 12.5 & \phn9.6 & 1.9 & 1.7 & 2600 & 1400 & 3.3 & 4600 & 2.1 \\
median  & 20 & 10.6 & \phn7.4 & 2.0 & 1.8 & 2100 & 1100 & 3.2 & 3500 & 1.9\\
\enddata
\tablenotetext{a}{The column density for the peak position of each core.}
\tablenotetext{b}{Beam-deconvolved radius of each core.}
\tablenotetext{c}{Number of associated IRAS point sources.}
\tablenotetext{d}{Number of associated MSX point sources.}
\tablenotetext{e}{The name of the associated H$^{13}$CO$^+$ cores.}
\end{deluxetable}
\clearpage

%%%%%%%%%%%%%%%%%%%%%%%%%%%%%%%%%%%%%%%%%%%%%%%%%%%%%%%%%%%%%%%%%%
%
% Table 4
%
%%%%%%%%%%%%%%%%%%%%%%%%%%%%%%%%%%%%%%%%%%%%%%%%%%%%%%%%%%%%%%%%%%
\begin{deluxetable}{lccclcccccc}
\tabletypesize{\scriptsize}
\rotate
\tablecaption{Average Physical Properties of C$^{18}$O Cores in Massive-Star-Forming Regions \label{table04}}
\tablewidth{0pt}
\tablehead{
\colhead{Region} & \colhead{Distance} & \colhead{Number of} & \colhead{Number of\tablenotemark{a}} & \colhead{$\Delta V$\tablenotemark{b}} & \colhead{$N$(H$_2$)} & \colhead{$M$} & \colhead{$r$\tablenotemark{c}} & \colhead{$n$(H$_2$)\tablenotemark{c}} & \colhead{$M_{\rm vir}$/$M$\tablenotemark{c,d}} & \colhead{Ref.}\\
 & \colhead{(pc)} & \colhead{C$^{18}$O Cores} & \colhead{Resolved Cores} & \colhead{(km s$^{-1}$)} & \colhead{(10$^{22}$ cm$^{-2}$)} & \colhead{($M_\sun$)} & \colhead{(pc)}  & \colhead{(10$^3$ cm$^{-3}$)} & 
}
\startdata
Orion B & \phn400 & 19 & 18 & 1.7 (1.7) & 2.1 (1.6)  & 250 (160) & 0.5 (0.5) & \phn7.6 (5.7) \tablenotemark{e} & 2.1 (1.4) \tablenotemark{e} & 1\\
Orion A & \phn480 & 19 & 12 & 1.7 (1.6) & 2.4 (1.7) & \nodata & \nodata & \nodata & \nodata & 2\\
Vela C & \phn700 & 13 & 10 & 2.5 (2.3) & 2.8 (2.8) & \nodata & \nodata & \nodata & \nodata & 3\\
Cepheus OB3 & \phn730 &\phn8 & \phn8 & 2.0 (1.9) \tablenotemark{f} & 1.5 (1.4) & \nodata & \nodata & \nodata & \nodata & 4\\
S35/37 & 1800 & 18 & \phn4 & 1.7 (1.6) & 1.7 (1.1) & 1100 (640) & 1.3 (1.3) & 11.0 (9.2) & 0.5 (0.6) & 5\\
$\eta$ Car GMC & 2500 & 15 & 11 & 3.2 (2.9) & 1.3 (1.1) & 2600 (2100) & 2.2 (2.0) & 1.2 (0.9) & 1.6 (1.3) & 6\\
Centaurus III & 3500 & 14 & 10 & 2.3 (2.3) & 1.3 (1.3) & 5400 (4700) & 2.8 (2.8) & 1.1 (1.0) \tablenotemark{e} & 0.8 (0.6) & 7\\
Centaurus II & 4500 & \phn5 & \phn5 & 3.1 (3.0) & 1.9 (1.9) & 20000 (24000) & 4.2 (3.8) & 1.1 (0.7) \tablenotemark{e} & 0.6 (0.5) & 7\\
Centaurus I & 5300 & 16 & 11 & 2.3 (2.2) & 1.3 (1.0) & 11000 (8600) & 4.0 (3.6) & 0.7 (0.7) \tablenotemark{e} & 0.5 (0.5) & 7\\
\enddata
\tablecomments{Median values are given in parentheses.}
\tablerefs{
(1)~Aoyama et al. 2001; (2)~Nagahama 1997; (3)~Yamaguchi et al. 1999b; (4)~Yu et al. 1996; (5)~Saito et al. 1999; (6)~this work; (7)~Saito et al. 2001.}
\tablenotetext{a}{Number of C$^{18}$O cores with $r_{\rm dec}$/$r$ $\gtrsim$ 0.8.}
\tablenotetext{b}{FWHM line width at the peak position of the core.}
\tablenotetext{c}{Values for resolved cores.}
\tablenotetext{d}{Calculated by using the equation ($M_{\rm vir}$/$M_\sun$) = 209 $\times$ ($r$/pc) ($\Delta V$/km s$^{-1}$)$^2$.}
\tablenotetext{e}{Calculated in this work.}
\tablenotetext{f}{Derived from the composite profile, i.e., $\Delta V_{\rm comp}$.}
\end{deluxetable}
\clearpage

%%%%%%%%%%%%%%%%%%%%%%%%%%%%%%%%%%%%%%%%%%%%%%%%%%%%%%%%%%%%%%%%%%
%
% Table 5
%
%%%%%%%%%%%%%%%%%%%%%%%%%%%%%%%%%%%%%%%%%%%%%%%%%%%%%%%%%%%%%%%%%%
\begin{deluxetable}{rrrrrrrrrrrccr}
\tabletypesize{\scriptsize}
\rotate
\tablecaption{Properties of the Associated IRAS Point Sources \label{table05}}
\tablewidth{0pt}
\tablehead{
& \multicolumn{4}{c}{Coordinate} & & \multicolumn{4}{c}{Flux Density}\\
\cline{2-5} \cline{7-10}
\colhead{Name} & \colhead{$l$} & \colhead{$b$} & \colhead{$\alpha$} & \colhead{$\delta$} & & $F_{12}$ & $F_{25}$ & $F_{60}$ & $F_{100}$ & \colhead{$L$} & \colhead{Quality} & \colhead{c.c.\tablenotemark{a}} & \colhead{C$^{18}$O\tablenotemark{b}}\\
& \colhead{($\arcdeg$)} & \colhead{($\arcdeg$)} & \colhead{(J2000)} & \colhead{(J2000)} & & \colhead{(Jy)} & \colhead{(Jy)} & \colhead{(Jy)} & \colhead{(Jy)} & \colhead{($L_\sun$)}
}
\startdata
10365$-$5803 & 286.203 & 0.170 & 10 38 30.6 & $-$58 19 00 & & 7.242 & 85.64 & 1173\phd\phn & 2782 & 30000 & 1333 & CABA & 5\\
10361$-$5830 & 286.375 & $-$0.255 & 10 38 03.8 & $-$58 46 18 & & 12.38\phn & 38.38 & 625.7 & 2156 & 21000 & 1333 & DCBD & 6\\
\enddata
\tablecomments{Units of right ascension are hours, minutes, and seconds, and units of declination are degrees, arcminutes, and arcseconds.}
\tablenotetext{a}{Correlation coefficient.}
\tablenotetext{b}{The name of the associated C$^{18}$O core.}
\end{deluxetable}
\clearpage

%%%%%%%%%%%%%%%%%%%%%%%%%%%%%%%%%%%%%%%%%%%%%%%%%%%%%%%%%%%%%%%%%%
%
% Table 6
%
%%%%%%%%%%%%%%%%%%%%%%%%%%%%%%%%%%%%%%%%%%%%%%%%%%%%%%%%%%%%%%%%%%
\begin{deluxetable}{clrrrrrrrrrccr}
\tabletypesize{\scriptsize}
\rotate
\tablecaption{Properties of the Associated MSX Point Sources \label{table06}}
\tablewidth{0pt}
\tablehead{
& & \multicolumn{4}{c}{Coordinate} & & \multicolumn{4}{c}{Flux Density}\\
\cline{3-6} \cline{8-11}
\colhead{No.} & \colhead{Name} & \colhead{$l$} & \colhead{$b$} & \colhead{$\alpha$} & \colhead{$\delta$} & & \colhead{Band-A} & \colhead{Band-C} & \colhead{Band-D} & \colhead{Band-E} & \colhead{Quality} & \colhead{C$^{18}$O\tablenotemark{a}}\\
& & \colhead{($\arcdeg$)} & \colhead{($\arcdeg$)} & \colhead{(J2000)} & \colhead{(J2000)} & & \colhead{(Jy)} & \colhead{(Jy)} & \colhead{(Jy)} & \colhead{(Jy)}
}
\startdata
1 & MSX6C G286.2086+00.1694 & 286.2086 & 0.1694 & 10 38 32.5 & $-$58 19 12 & & 1.3532\phn & 2.8816 & 7.1817 & 40.572\phn & 4444 & 5\\
2 & MSX6C G286.3579-00.2933 & 286.3579 & $-$0.2933 & 10 37 48.4 & $-$58 47 48 & & 0.71259 & 1.8153 & 2.6774 & 6.0653 & 4344 & 6\\
3 & MSX6C G286.3747-00.2630 & 286.3747 & $-$0.2630 & 10 38 02.0 & $-$58 46 43 & & 3.5907\phn & 4.7559 & 2.4093 & 7.5775 & 4444 & 6\\
4 & MSX6C G286.3773-00.2563 & 286.3773 & $-$0.2563 & 10 38 04.6 & $-$58 46 26 & & 1.6278\phn & 2.9183 & 3.8554 & 12.095\phn & 4444 & 6\\
5 & MSX6C G287.2238-00.5339 & 287.2238 & $-$0.5339 & 10 42 48.9 & $-$59 25 29 & & 1.2040\phn & 1.6909 & 1.5151 & 4.3560 & 4344 & 9\\
6 & MSX6C G287.3399-00.6008 & 287.3399 & $-$0.6008 & 10 43 22.3 & $-$59 32 19 & & 0.80028 & 2.9386 & 5.9723 & 12.516\phn & 4444 & 10\\
7 & MSX6C G287.3506-00.5900 & 287.3506 & $-$0.5900 & 10 43 29.2 & $-$59 32 03 & & 0.55922 & 2.0313 & 3.9602 & 7.9096 & 4444 & 10\\
8 & MSX6C G287.6393-00.7219 & 287.6393 & $-$0.7219 & 10 45 00.9 & $-$59 47 10 & & 3.4863\phn & 4.2647 & 3.6021 & 17.835\phn & 4444 & 12\\
\enddata
\tablecomments{Units of right ascension are hours, minutes, and seconds, and units of declination are degrees, arcminutes, and arcseconds.}
\tablenotetext{a}{The name of the associated C$^{18}$O core.}
\end{deluxetable}
\clearpage

%%%%%%%%%%%%%%%%%%%%%%%%%%%%%%%%%%%%%%%%%%%%%%%%%%%%%%%%%%%%%%%%%%
%
% Table 7
%
%%%%%%%%%%%%%%%%%%%%%%%%%%%%%%%%%%%%%%%%%%%%%%%%%%%%%%%%%%%%%%%%%%
\begin{deluxetable}{cccccccccccrrccc}
\tabletypesize{\scriptsize}
\rotate
\tablecaption{Physical Properties of H$^{13}$CO$^+$ Cores \label{table07}}
\tablewidth{0pt}
\tablehead{
\colhead{Name}
 & \colhead{$\alpha$}
 & \colhead{$\delta$}
 & \colhead{$T_{\rm R}^*$}
 & \colhead{$V_{\rm LSR}$}
 & \colhead{$\Delta V$}
 & \colhead{$T_{\rm ex}$}
 & \colhead{$\tau$}
 & \colhead{$N$(H$_2$)\tablenotemark{a}}
 & \colhead{$r$}
 & \colhead{$\Delta V_{\rm comp}$}
 & \colhead{$M$}
 & \colhead{$M_{\rm vir}$}
 & \colhead{$\langle$$N$(H$_2$)$\rangle$\tablenotemark{a}}
 & \colhead{$n$(H$_2$)\tablenotemark{b}}
 & \colhead{C$^{18}$O\tablenotemark{c}}\\
 & \colhead{(J2000)}
 & \colhead{(J2000)}
 & \colhead{(K)}
 & \colhead{(km s$^{-1}$)}
 & \colhead{(km s$^{-1}$)}
 & \colhead{(K)}
 &
 &
 & \colhead{(pc)}
 & \colhead{(km s$^{-1}$)}
 & \colhead{($M_\sun$)}
 & \colhead{($M_\sun$)}
 &
 &
}
\startdata
1 & 10 38 31.7 & $-$58 18 44 & 0.38 & $-$19.4 & 2.9 & 30 & 0.018 & 4.5 & 0.77 & 3.8 & 1400 & 2300 & 3.4 & 1.1 & \phn5\\
2 & 10 37 52.9 & $-$58 47 20 & 0.41 & $-$22.5 & 3.0 & 55 & 0.011 & 7.6 & 0.27 & 3.0 & 400 & 500 & 7.6 & 6.8 & \phn6\\
3 & 10 42 47.8 & $-$59 26 24 & 0.35 & $-$17.3 & 1.3 & 35 & 0.017 & 2.0 & \nodata & \nodata & \nodata & \nodata & \nodata & \nodata & \phn9\\
4 & 10 47 46.7 & $-$60 26 26 & 0.81 & $-$17.8 & 1.6 & 30 & 0.035 & 5.0 & 0.39 & 1.6 & 470 & 200 & 4.4 & 2.8 & 15\\
\enddata
\tablecomments{Units of right ascension are hours, minutes, and seconds, and units of declination are degrees, arcminutes, and arcseconds.}
\tablenotetext{a}{Units are in 10$^{22}$ cm$^{-2}$.}
\tablenotetext{b}{Units are in 10$^4$ cm$^{-3}$.}
\tablenotetext{c}{The name of the associated C$^{18}$O core.}
\end{deluxetable}
\clearpage

%%%%%%%%%%%%%%%%%%%%%%%%%%%%%%%%%%%%%%%%%%%%%%%%%%%%%%%%%%%%%%%%%%
%
% Table 8
%
%%%%%%%%%%%%%%%%%%%%%%%%%%%%%%%%%%%%%%%%%%%%%%%%%%%%%%%%%%%%%%%%%%
\begin{deluxetable}{lccccccccc}
\tabletypesize{\scriptsize}
\rotate
\tablecaption{Properties of the Outflow \label{table08}}
\tablewidth{0pt}
\tablehead{
 & & & & & & \colhead{Kinetic} & \colhead{Mechanical} & & \colhead{Dynamical}\\
 & \colhead{$V_{\rm range}$\tablenotemark{a}} & \colhead{$V_{\rm char}$\tablenotemark{b}} & \colhead{Size\tablenotemark{c}} & \colhead{Radius} & \colhead{Mass} & \colhead{Energy} & \colhead{Luminosity} & \colhead{Momentum} & \colhead{Timescale}\\
\colhead{Component} & \colhead{(km s$^{-1}$)} & \colhead{(km s$^{-1}$)} & \colhead{(pc)} & \colhead{(pc)} & \colhead{($M_\sun$)} & \colhead{(10$^{45}$ ergs)} & \colhead{($L_\sun$)} & \colhead{($M_\sun$ km s$^{-1}$)} & \colhead{(10$^4$ yr)}
}
\startdata
Blue & ($-$35, $-$25) & 16.5 & 0.41 & 0.41 & 1.2 & 1.7 & 0.39 & 14 & 2.4\\
Red & ($-$15, \phn$-$5) & 13.5 & 0.41 & 0.55 & 2.1 & 1.9 & 0.37 & 20 & 4.0\\
\enddata
\tablenotetext{a}{Velocity range where the high velocity emission is detected at $>$ 1-$\sigma$ noise level ($T_{\rm R}^*$$\sim$0.8 K).}
\tablenotetext{b}{Maximum velocity shift of the $^{12}$CO ($J$ = 2--1) emission from $V_{\rm LSR}$ = $-$18.5 km s$^{-1}$.}
\tablenotetext{c}{The maximum separation of the outflow lobes from the position of the driving source.}
\end{deluxetable}
\clearpage

%%%%%%%%%%%%%%%%%%%%%%%%%%%%%%%%%%%%%%%%%%%%%%%%%%%%%%%%%%%%%%%%%%
%
% Table 9
%
%%%%%%%%%%%%%%%%%%%%%%%%%%%%%%%%%%%%%%%%%%%%%%%%%%%%%%%%%%%%%%%%%%
\begin{deluxetable}{lclcccccc}
\tabletypesize{\scriptsize}
\rotate
\tablecaption{Average Physical Properties of C$^{18}$O Cores in the $\eta$ Car GMC\label{table09}}
\tablewidth{0pt}
\tablehead{
 & \colhead{Group\tablenotemark{a}} & \colhead{Name\tablenotemark{b}} & \colhead{$\Delta V_{\rm comp}$} & \colhead{$N$(H$_2$)\tablenotemark{c}} & \colhead{$M$} & \colhead{$r$\tablenotemark{d}} & \colhead{$n$(H$_2$)\tablenotemark{d}} & \colhead{$M_{\rm vir}$/$M$\tablenotemark{d}}\\
 & & & \colhead{(km s$^{-1}$)} & & \colhead{($M_\sun$)} & \colhead{(pc)} & \colhead{($\times$ 10$^3$ cm$^{-3}$)}
}
\startdata
Cores with IRAS point sources& A & 5, 6 & 3.2 (3.2) & 26.1 (26.1) & 4900 (4900) & 1.9 (1.9) & 2.5 (2.5) & 1.0 (1.0)\\
Other star-forming cores & B & 9, (10), 12, 15 & 3.2 (3.2) & 9.9 (9.8) & 1600 (1700) & 2.0 (2.0) & 0.9 (0.9) & 1.9 (1.8)\\
All star-forming cores & A+B & 5, 6, 9, (10), 12, 15 & 3.2 (3.2) & 15.3 (10.5) & 2700 (2300) & 2.0 (2.0) & 1.5 (1.1) & 1.5 (1.6)\\
Non-star-forming cores & C & 1, 2, 3, (4), 7, 8, (11), 13, (14) & 3.3 (3.2) & 10.7 (10.6) & 2600 (2100) & 2.4 (2.4) & 0.9 (0.8) & 1.8 (1.7)\\
All C$^{18}$O cores & \nodata & all & 3.3 (3.2) & 12.5 (10.6) & 2600 (2100) & 2.2 (2.0) & 1.2 (0.9) & 1.7 (1.6)\\
\enddata
\tablecomments{Median values are given in parentheses.}
\tablenotetext{a}{See text (\S~4.1).}
\tablenotetext{b}{The Names of the associated C$^{18}$O cores. The names of the unresolved cores are given in parentheses.}
\tablenotetext{c}{Units are in 10$^{21}$ cm$^{-2}$.}
\tablenotetext{d}{Values for resolved cores.}
\end{deluxetable}

\begin{thebibliography}{}
\bibitem[Allen(1972)]{all72} Allen, C. W. 1972, Astronomical Quantities (London: Athlone)
%
\bibitem[Alves \& Homeier(2003)]{alv03} Alves, J. \& Homeier, N. 2003, \apjl, 589, L45
%
\bibitem[Andr\'{e}, Word-Thompson, \& Barsony(1993)]{and93} Andr\'{e}, P., Word-Thompson, D., \& Barsony, M. 1993 \apj, 406, 122
%
\bibitem[Aoyama et al.(2001)]{aoy01} Aoyama, H., Mizuno, N., Yamamoto, H., Onishi, T., Mizuno, A., \& Fukui, Y. 2001, \pasj, 53, 1053
%
\bibitem[Bergin et al.(1997)]{ber97} Bergin, E. A, Goldsmith, P. F., Snell, R. L., \& Langer, W. D. 1997, \apj, 482, 285
%
\bibitem[Bertoldi \& McKee(1992)]{ber92} Bertoldi, F. \& McKee, C. F. 1992, \apj, 395, 140
%
\bibitem[Bertsch et al.(1993)]{ber93} Bertsch, D. L., Dame, T. M., Fichtel, C. E., Hunter, S. D., Sreekumar, P., Stacy, J. G., \& Thaddeus, P. 1993, \apj, 416, 587
%
\bibitem[Beuther et al.(2002a)]{beu02a} Beuther, H., Schilke, P., Menten, K. M., Motte, F., Sridharan, T. K., \& Wyrowski, F. 2002, \apj, 566, 945
%
\bibitem[Beuther et al.(2002b)]{beu02b} Beuther, H., Schilke, P., Sridharan, T. K., Menten, C. M. Walmsley, \& Wyrowski, F. 2002, \aap, 383, 892
%
\bibitem[Bloemen et al.(1986)]{blo86} Bloemen J. B. G. M., et al. 1986, \aap, 154, 25
%
\bibitem[van Boekel et al.(2003)]{boe03} van Boekel, R., et al. 2003, \aap, 410, L37
%
\bibitem[Bonnell, Bate, \& Zinnecker(1998)]{bon98} Bonnell, I. A., Bate, M. R., \& Zinnecker, H. 1998, \mnras, 298, 93
%
\bibitem[Bronfman, Nyman, \& May(1996)]{bro96} Bronfman, L., Nyman, L. -\AA.,  \& May, J. 1996, \aaps, 115, 81
%
\bibitem[Brooks(2000)]{bro00} Brooks, K. 2000, PhD thesis, Univ. New South Wales
%
\bibitem[Brooks et al.(2003)]{bro03} Brooks, K. J., Cox, P., Schneider, N., Storey, J. W. V., Poglitsch, A., Geis, N., \& Bronfman, L. 2003, \aap, 412, 751
%
\bibitem[Brooks, Whiteoak, \& Storey(1998)]{bro98} Brooks, K. J., Whiteoak, J. B., \& Storey, J. W. V. 1998, \pasa, 15, 202
%
\bibitem[Cabrit \& Bertout(1992)]{cab92} Cabrit, S. \& Bertout, C. 1992, \aap, 261, 274
%
\bibitem[Caselli \& Myers(1995)]{cas95} Casseli, P. \& Myers, P. C. 1995, \apj, 446, 665
%
\bibitem[Castor, McCray, \& Weaver(1975)]{cas75} Castor, J., McCray, R., \& Weaver, R. 1975, \apj, 200, 107
%
\bibitem[Cox \& Bronfman(1995)]{cox95} Cox, P. \& Bronfman, L. 1995, \aap, 299, 583
%
\bibitem[Davidson \& Humphreys(1997)]{dav97} Davidson, K. \& Humphreys, R. M. 1997, \araa, 35, 1
%
\bibitem[de Graauw et al.(1981)]{deg81} de Graauw, T., Lidholm, S., Fitton, B., Beckman, J., Israel, F. P., Nieuwenhuijzen, H., \& Vermue, J. 1981, \aap, 102, 257
%
\bibitem[Dickman(1978)]{dic78} Dickman, R. L. 1978, \apjs, 37, 407
%
\bibitem[Dobashi, Bernard, \& Fukui(1996)]{dob96} Dobashi, K., Bernard, J.~P., \& Fukui, Y. 1996, \apj, 466, 282
%
\bibitem[Dobashi et al.(1994)]{dob94} Dobashi, K., Bernard, J.~P., Yonekura, Y., \& Fukui, Y. 1994, \apjs, 95, 419
%
\bibitem[Drissen et al.(1995)]{dri95} Drissen, L., Moffat, A. F. J., Walborn, N. R., \& Shara, M. M. 1995, \aj, 110, 2235
%
\bibitem[Elmegreen(1993)]{elm93} Elmegreen, B. G. 1993, , in Protostars and Planets III, ed. E. H. Levy \& J. I. Lunine (Tucson: Univ. Arizone Press), 97
%
\bibitem[Evans(1999)]{eva99} Evans, N. J., II. 1999, \araa, 37, 311
%
\bibitem[Evans et al.(2002)]{eva02} Evans, N. J., II, Shirley, Y. L., Mueller, K. E., \& Knez, C. 2002, in ASP Conf. Ser. 267, The Earliest Stages of Massive Star Birth, ed. P.A. Crowther (San Francisco: ASP), 17
%
\bibitem[Ezawa et al.(2004)]{eza04} Ezawa, H., Kawabe, R., Kohno, K., and Yamamoto, S. 2004, Proc.\ SPIE, 5489, 763
%
\bibitem[Fa\'{u}ndez et al.(2004)]{fau04} Fa\'{u}ndez, S., Bronfman, L., Garay, G., Chini, R., May, J., \& Nyman, L. \AA. 2004, \aap, in press
%
\bibitem[Feinstein(1995)]{fei95} Feinstein, A. 1995, \rmxaa, 2, 57
%
\bibitem[Forbrich et al.(2004)]{for04} Forbrich, J., Schreyer, K., Posselt, B., Klein, R., \& Henning, Th. 2004, \apj, 602, 843
%
\bibitem[Frerking, Langer, \& Wilson(1982)]{fre82} Frerking, M.~A., Langer, W.~D., \& Wilson, R.~W. 1982, \apj, 262, 590
%
\bibitem[Fukui(1989)]{fuk89} Fukui, Y. 1989, Proc. ESO Workshop on Low Mass Star Formation and Pre-Main Sequence Objects, ed. B. Reipurth (ESO: Garching), 95
%
\bibitem[Fukui et al.(1993)]{fuk93} Fukui, Y., Iwata, T., Mizuno, A., Bally, J., \& Lane, A. P. 1993, in Protostars and Planets III, ed. E. H. Levy \& J. I. Lunine (Tucson: Univ. Arizone Press), 603
%
\bibitem[Fukui et al.(1991)]{fuk91} Fukui, Y., Ogawa, H., Kawabata, K., Mizuno, A., \& Sugitani, K. 1991, in IAU Symp. 148, The Magellanic Clouds, ed. R. Haynes \& D. Mline (Dordrecht: Kluwer), 105
%
\bibitem[Fukui et al.(1999)]{fuk99} Fukui, Y., Onishi, T., Abe, R., Kawamura, A., Tachihara, K., Yamaguchi, R., Mizuno, A., \& Ogawa, H. 1999, \pasj, 51, 751
%
\bibitem[Fukui \& Sakakibara(1992)]{fuk92} Fukui, Y. \& Sakakibara, O. 1992, Mitsubishi Electric ADVANCE, 60, 11
%
\bibitem[Fukui \& Yonekura(1998)]{fuk98} Fukui, Y. \& Yonekura, Y. 1998, in IAU Symp. 179, New Horizons from Multi-Wavelength Sky Surveys, ed. B. J. McLean, D. A. Golombek, J. J. E. Hayes, \& H. E. Payne (Dordrecht: Kluwer), 165
%
\bibitem[Garay et al.(2004)]{gar04} Garay, G., Fa\'{u}ndez, S., Mardones, D., Bronfman, L., Chini, R., \& Nyman, L. -\AA. 2004, \apj, 610, 313
%
\bibitem[Gaustad et al.(2001)]{gau01} Gaustad, J. E., McCullough, P. R., Rosing, W., \& Van Buren, D. 2001, \pasp, 113, 1326
%
\bibitem[Goldsmith and Langer(1978)]{gol78} Goldsmith, P. F. \& Langer W. D. 1978, \apj, 222, 881
%
\bibitem[Goodman et al.(1998)]{goo98} Goodman, A. A., Barranco, J. A., Wilner, D. J., \& Heyer, M. H. 1998, \apj, 504, 223
%
\bibitem[Grabelsky et al.(1988)]{gra88} Grabelsky, D. A., Cohen, R. S., Bronfman, L., \& Thaddeus, P. 1988, \apj, 331, 181
%
\bibitem[Grabelsky et al.(1987)]{gra87} Grabelsky, D. A., Cohen, R. S., Bronfman, L., Thaddeus, P., \& May, J. 1987, \apj, 315, 122
%
\bibitem[Hara et al.(1999)]{har99} Hara, A., Tachihara, K., Mizuno, A., Onishi, T., Kawamura, A., Obayashi, A., \& Fukui, Y. 1999, \pasj, 51, 895
%
\bibitem[Harju, Walmsley, \& Wouterloot(1993)]{har93} Harju, J., Walmsley, C. M., \& Wouterloot, J. G. A. 1993, \aaps, 98, 51
%
\bibitem[van der Hucht(2001)]{huc01} van der Hucht, K. A. 2001, \nar, 45, 135
%
\bibitem[Hunter et al.(1997)]{hun97} Hunter, S. D., et al. 1997, \apj, 481, 205
%
\bibitem[Jijina, Myers, \& Adams(1999)]{jij99} Jijina, J., Myers, P. C., \& Adams, F. C. 1999, \apjs, 125, 161
%
\bibitem[Johansson et al.(1984)]{joh84} Johansson, L. E. B., Andersson, C., Elld\'er, P., Friberg, P., Hjalmarson, \AA., H\"oglund, B., Irvine, W. M., Olofsson, H., \& Rydbeck, G. 1984, \aap, 130, 227
%
\bibitem[Jones(1973)]{jon73} Jones, B. B. 1973, Aust. J. Phys., 26, 545
%
\bibitem[J\o rgensen, Sch\"{o}ier, \& van Dishoeck(2004)]{jor04} J\o rgensen, J. K., Sch\"{o}ier, F. L., \& van Dishoeck, E. F. 2004, \aap, 416, 603
%
\bibitem[Juvela(1996)]{juv96} Juvela, M. 1996, \aaps, 118, 191
%
\bibitem[Kato et al.(1999)]{kat99} Kato, S., Mizuno, N., Asayama, S., Mizuno, A., Ogawa, H., \& Fukui, Y.  1999, \pasj, 51, 883
%
\bibitem[Kawamura et al.(1998)]{kaw98} Kawamura, A., Onishi, T., Yonekura, Y., Dobashi, K., Mizuno, A., Ogawa, H., \& Fukui Y. 1998, \apjs, 117, 387
%
\bibitem[Kim et al.(2004)]{kim04} Kim, B. G., Kawamura, A., Yonekura, Y., \& Fukui, Y. 2004, \pasj, 56, 313
%
\bibitem[Kohno et al.(2003)]{koh03} Kohno, K., et al. 2003, in Springer Proceedings in Physics 91, The Dense Interstellar Medium in Galaxies, ed. S. Pfalzner, C. Kramer, C. Staubmeier, \& A. Heithausen (Berlin: Springer), 349
%
\bibitem[Koyama \& Inutsuka(2000)]{koy00} Koyama, H., \& Inutsuka, S. 2000, \apj, 532, 980
%
\bibitem[Koyama \& Inutsuka(2002)]{koy02} Koyama, H., \& Inutsuka, S. 2002, \apjl, 564, L97
%
\bibitem[Kurtz et al.(2000)]{kur00} Kurtz, S., Cesaroni, R., Churchwell, E., Hofner, P., \& Walmsley, C. M. 2000, Protostars and Planets IV, ed. V. Mannings, A. Boss, \& S. Russel (Tucson: Univ. of Arizona Press), 299
%
\bibitem[Kutner \& Ulich(1981)]{kut81} Kutner, M. L. \& Ulich B. L. 1981, \apj, 250, 341
%
\bibitem[Lada, Bally, \& Stark(1991)]{lad91} Lada, E. A., Bally, J., \& Stark, A. A. 1991, \apj, 368, 432
%
\bibitem[Ladd, Myers, \& Goodman(1994)]{lad94} Ladd, E. F., Myers, P. C., \& Goodman, A. A. 1994, \apj, 433, 117
%
\bibitem[Ladd et al.(2005)]{lad05} Ladd, N., Purcell, C., Wong, T., \& Robertson, S. 2005, \pasa, 22, 62
%
\bibitem[Levreault(1988)]{lev88} Levreault, R. M. 1988, \apjs, 67, 283
%
\bibitem[McKee \& Tan(2003)]{mck03} McKee, C. F. \& Tan, J. C. 2003, \apj, 585, 850
%
\bibitem[Ma\'{i}z-Apell\'{a}niz et al.(2004)]{mai04} Ma\'{i}z-Apell\'{a}niz, J., Walborn, N. R., Galu\'{e}, H. \'{A}., \& Wei, L. H. 2004, \apjs, 151, 103
%
\bibitem[Markova et al.(2004)]{mar04} Markova, N., Puls, J., Repolust, T., \& Markov, H. 2004, \aap, 413, 693
%
\bibitem[Megeath et al.(1996)]{meg96} Megeath, S. T., Cox, P., Bronfman, L., \& Roelfsema, P. R. 1996, \aap, 305, 296
%
\bibitem[Mizuno et al.(1998)]{miz98} Mizuno, A., et al. 1998, \apjl, 507, L83
%
\bibitem[Mizuno et al.(1994)]{miz94} Mizuno, A., Onishi, T., Hayashi, M., Ohashi, N., Sunada, K., Hasegawa, T., \& Fukui, Y. 1994, \nat, 368, 719
%
\bibitem[Mizuno et al.(1995)]{miz95} Mizuno, A., Onishi, T., Yonekura, Y., Nagahama, T., Ogawa, H., \& Fukui, Y. 1995, \apjl, 445, L161
%
\bibitem[Mizuno et al.(1999)]{miz99} Mizuno, A., et al. 1999, \pasj, 51, 859
%
\bibitem[Mueller et al.(2002)]{mue02} Mueller, K. E., Shirley, Y. L., Evans II, N. J., \& Jacobson, H. R. 2002, \apjs, 143, 469
%
\bibitem[Myers, Ladd, \& Fuller(1991)]{mye91} Myers, P. C., Ladd, E. F., \& Fuller, G. A. 1991, \apjl, 372, L95
%
\bibitem[Nagahama(1997)]{nag97} Nagahama T. 1997, PhD thesis, Nagoya University
%
\bibitem[Nakano(1998)]{nak98} Nakano, T. 1998, \apj, 494, 587
%
\bibitem[Nozawa et al.(1991)]{noz91} Nozawa, S., Mizuno, A., Teshima, Y., Ogawa, H., \& Fukui, Y. 1991, \apjs, 77, 647
%
\bibitem[Obayashi et al.(1998)]{oba98} Obayashi, A., Kun, M., Sato, F., Yonekura, Y., \& Fukui, Y. 1998, \aj, 115, 274
%
\bibitem[Ogawa et al.(1990)]{oga90} Ogawa, H., Mizuno, A., Hoko, H., Ishikawa, H., \& Fukui, Y. 1990, Int. J. Infrared Millimeter Waves, 11, 717
%
\bibitem[Onishi et al.(1996)]{oni96} Onishi, T., Mizuno, A., Kawamura, A., Ogawa, H., \& Fukui, Y. 1996, \apj, 465, 815
%
\bibitem[Onishi et al.(1998)]{oni98} Onishi, T., Mizuno, A., Kawamura, A., Ogawa, H., \& Fukui, Y. 1998, \apj, 502, 296
%
\bibitem[Onishi et al.(2002)]{oni02} Onishi, T., Mizuno, A., Kawamura, A., Tachihara, K., \& Fukui, Y. 2002, \apj, 575, 950
%
\bibitem[Onishi et al.(1999)]{oni99} Onishi, T., et al.\ 1999, \pasj, 51, 871
%
\bibitem[Plume, Jaffe, \& Evans(1992)]{plu92} Plume, R., Jaffe, D. T., \& Evans II, E. J. 1992, \apjs, 78, 505
%
\bibitem[Plume et al.(1997)]{plu97} Plume, R., Jaffe, D. T., Evans II, E. J., Mart\'in-Pintado, J., \& G\'omez-Gonz\'alez, J. 1997, \apj, 476, 730
%
\bibitem[Rathborne et al.(2004)]{rat04} Rathborne, J. M., Brooks, K. J., Burton, M. G., Cohen, M., \& Bontemps, S. 2004, \aap, 418, 563
%
\bibitem[Saito et al.(2001)]{sai01} Saito, H., Mizuno, N., Moriguchi, Y., Matsunaga, K., Onishi, T., Mizuno, A., \& Fukui, Y. 2001, \pasj, 53, 1037
%
\bibitem[Saito et al.(1999)]{sai99} Saito, H., Tachihara, K., Onishi, T., Yamaguchi, N., Mizuno, N., Mizuno, A., Ogawa, H., \& Fukui, Y. 1999, \pasj, 51, 819
%
\bibitem[Scalo(1986)]{sca86} Scalo, J. M. 1986, Fund. Cosmic Phys., 11, 1
%
\bibitem[Sekimoto et al.(2001)]{sek01} Sekimoto, Y., et al. 2001, \pasj, 53, 951
%
\bibitem[Seward et al.(1979)]{sew79} Seward, F. D., Forman, W. R., Giacconi, R., Griffiths, R. E., Harnden Jr., F. R., Jones, C., \& Pye, J. P. 1979, \apjl, 234, L55
%
\bibitem[Shepherd \& Churchwell(1996)]{she96} Shepherd, D.S. \& Churchwell, E. 1996, \apj, 472, 225
%
\bibitem[Shirley et al.(2003)]{shi03} Shirley, Y. L., Evans II, N. J., Young, K. E., Knez, C., \& Jaffe, D. T. 2003, \apjs, 149, 375
%
\bibitem[Shu, Adams, \& Lizano(1987)]{shu87} Shu, F. H., Adams, F. C., \& Lizano, S. 1987, \araa, 25, 23
%
\bibitem[Smith et al.(2003)]{smi03} Smith, N., Davidson, K., Gull, T. R., Ishibashi, K., \& Hillier, D. J. 2003, \apj, 586, 432
%
\bibitem[Smith et al.(2000)]{smi00} Smith, N., Egan, M. P., Carey, S., Price, S. D., Morse, J. A., \& Price, P. A. 2000, \apjl, 532, L145
%
\bibitem[Solomon et al.(1987)]{sol87} Solomon, P. M., Rivolo, A. R., Barrett, J., \& Yahil, A. 1987, \apj, 319, 730
%
\bibitem[Sorai et al.(2000)]{sor00} Sorai, K., Sunada, K., Okumura, S. K., Iwasa, T., Tanaka, A., Natori, N., \& Onuki, H. 2000, Proc.\ SPIE, 4015, 86
%
\bibitem[Sridharan et al.(2002)]{sri02} Sridharan, T. K., Beuther, H., Schilke, P., Menten, K. M., \& Wyrowski, F. 2002, \apj, 566, 931
%
\bibitem[Stacy et al.(1988)]{sta88} Stacy, J. G., Benson, P. J., Myers, P. C., \& Goodman, A. A. 1988, in Interstellar Matter: Proc. 2d Haystack Obs.\ Meeting, ed. J. M. Moran \& P. T. P. Ho (New York: Gordon \& Breach), 179
%
\bibitem[Strong et al.(1988)]{str88} Strong, A. W., et al. 1988, \aap, 207, 1
%
\bibitem[Sugitani et al.(1989)]{sug89} Sugitani, K., Fukui, Y., Mizuno, A., \& Ohashi, N. 1989, \apjl, 342, L87
%
\bibitem[Sutton et al.(1985)]{sut85} Sutton, E. C., Blake, G. A., Masson, C. R., \& Phillips, T. G. 1985, \apjs, 58, 341
%
\bibitem[Tachihara, Mizuno, \& Fukui(2000)]{tac00} Tachihara, K., Mizuno, A., \& Fukui, Y. 2000, \apj, 528, 817
%
\bibitem[Tachihara et al.(2002)]{tac02} Tachihara, K., Onishi, T., Mizuno, A., \& Fukui, Y. 2002, \aap, 385, 909
%
\bibitem[Tateyama, Strauss, \& Kaufmann(1991)]{tat91} Tateyama, C. E., Strauss, F. M., \& Kaufmann, P. 1991, \mnras, 249, 716
%
\bibitem[Ulich \& Haas(1976)]{uli76} Ulich, B. L. \& Haas, R. W. 1976, \apjs, 30, 247
%
\bibitem[V\'{a}zquez-Semadeni et al.(2000)]{vaz00} V\'{a}zquez-Semadeni, E., Ostriker, E. C., Passot, T., Gammie, C. F., \& Stone, J.M. 2000, Protostars and Planets IV, ed. V. Mannings, A. Boss, \& S. Russel (Tucson: Univ. of Arizona Press), 3
%
\bibitem[Warin, Benayoun, \& Viala(1996)]{war96} Warin, S., Benayoun, J. J., \& Viala, Y. P. 1996, \aap, 308, 535
%
\bibitem[Weaver et al.(1977)]{wea77} Weaver, R., McCray, R., Castor, J., Shapiro, P., \& Moore, R. 1977, \apj, s218, 377
%
\bibitem[Whiteoak(1994)]{whi94} Whiteoak, J. B. Z. 1994, \apj, 429, 225
%
\bibitem[Wouterloot, Walmsley, \& Henkel(1988)]{wou88} Wouterloot, J. G. A., Walmsley, C. M., \& Henkel, C. 1988, \aap, 203, 367
%
\bibitem[Wu et al.(2004)]{wu04} Wu, Y., Wei, Y., Zhao, M., Shi, Y., Yu, W., Qin, S., \& Huang, M. 2004, \aap, 426, 503
%
\bibitem[Yamaguchi et al.(1999a)]{yam99} Yamaguchi, N., Mizuno, N., Saito, H., Matsunaga, K., Mizuno, A., Ogawa, H., \& Fukui, Y. 1999a, \pasj, 51, 775
%
\bibitem[Yamaguchi et al.(1999b)]{yam99b} Yamaguchi, R., Saito, H., Mizuno, N., Mine, Y., Mizuno, A., Ogawa, H., \& Fukui, Y. 1999b, \pasj, 51, 791
%
\bibitem[Yonekura et al.(1998)]{yon98} Yonekura, Y., Dobashi, K., Hayashi, Y., Sato, F., Ogawa, H., \& Fukui, Y. 1998, \aj, 115, 2009
%
\bibitem[Yonekura et al.(1997)]{yon97} Yonekura, Y., Dobashi, K., Mizuno, A., Ogawa, H., \& Fukui, Y. 1997, \apjs, 110, 21
%
\bibitem[Yonekura et al.(1999)]{yon99} Yonekura, Y., Mizuno, N., Saito, H., Mizuno, A., Ogawa, H., \& Fukui, Y. 1999, \pasj, 51, 911
%
\bibitem[Yu, Nagahama, \& Fukui(1996)]{yu96} Yu, Z.-A., Nagahama, T., \& Fukui, Y. 1996, \apj, 471, 867
%
\bibitem[Zhang et al.(2001a)]{zha01a} Zhang, Q., Hunter, T.R., Brand, J., Sridharan, T.K., Molinari, S., Kramer, M.A., \& Cesaroni, R. 2001a, \apjl, 552, L167
%
\bibitem[Zhang et al.(2001b)]{zha01b} Zhang, X., Lee, Y., Bolatto, A., \& Stark, A. A. 2001b, \apj, 553, 274
%
\bibitem[Zinchenko, Mattila, \& Toriseva(1995)]{zin95} Zinchenko, I., Mattila, K., \& Toriseva, M. 1995, \aaps, 111, 95
%
\bibitem[Zinchenko, Pirogov, \& Toriseva(1998)]{zin98} Zinchenko, I., Pirogov, L., \& Toriseva, M. 1998, \aaps, 133, 337
%
\end{thebibliography}
\end{document}